\documentclass{article}
\usepackage{hyperref}
\usepackage{amsmath}
\usepackage{amsfonts}
\renewcommand{\theequation}{\arabic{section}.\arabic{equation}}

\begin{document}

\title{Large-size expansion for triangular Wilson loops in confining gauge theories}
\author{P. V. Pobylitsa\\ \emph{Petersburg Nuclear Physics Institute}\\ \emph{Gatchina, 188300, St. Petersburg, Russia}\footnote{On leave of absence}}
\date{}
\maketitle

\textbf{Abstract}

The asymptotic behavior of Wilson loops in the large-size limit ($L\rightarrow
\infty$) in confining gauge theories with area law is controlled by effective
string theory (EST). The $L^{-2}$ term of the large-size expansion for the
logarithm of Wilson loop appears within EST as a 2-loop correction.
Ultraviolet divergences of this 2-loop correction for polygonal contours can
be renormalized using an analytical regularization constructed in terms of
Schwarz-Christoffel mapping. In the case of triangular Wilson loops this
method leads to a simple final expression for the $L^{-2}$ term.

\tableofcontents

\newpage

\section{Introduction}

\setcounter{equation}{0} 

\subsection{Large-size expansion for Wilson loops}

Gauge theories obeying area law \cite{Wilson-74}
\begin{equation}
\lim_{\left|  C\right|  \rightarrow\infty}\frac{1}{S\left(  C\right)  }\ln
W\left(  C\right)  =-\sigma\neq0 \label{Wilson-area-law}
\end{equation}
for Wilson loops
\begin{equation}
W\left(  C\right)  =\left\langle \mathrm{Tr}\,\mathrm{P}\,\exp\left[  i\oint_{C}dx^{\mu
}A_{\mu}\left(  x\right)  \right]  \right\rangle
\end{equation}
(with large flat contour $C$ bounding a region with area $S\left(  C\right)
$) play an important role as models of the heavy-quark confinement and are
often briefly called confining gauge theories (although the problem of
confinement in real QCD is more complicated). In our modern understanding of
confining gauge theories, the area law is considered as a secondary
manifestation of a much deeper phenomenon of \emph{effective string
formation}. This phenomenon is described by \emph{effective string theory}
(EST) which provides a detailed information about the asymptotic behavior of
Wilson loops in the limit of large size of contour $C$ (denoted informally as
$\left|  C\right|  \rightarrow\infty$ in eq.~(\ref{Wilson-area-law})). EST
assumes that the asymptotic behavior of Wilson loops in the limit of large
size of contour $C$ can be described by a functional integral over surfaces
$\Sigma$ bounded by contour $C$:
\begin{equation}
W\left(  C\right)  \rightarrow\mathrm{const}\int_{\partial\Sigma=C}D\Sigma
\exp\left(  -\mathcal{S}_{\mathrm{EST}}\left[  \Sigma\right]  \right)  .
\label{W-C-EST}
\end{equation}
The idea that Wilson loops $W\left(  C\right)  $ can be approximated by
functional integral (\ref{W-C-EST}) has a long history. Starting from
qualitative and heuristic arguments \cite{Polyakov-1980}, one can try to
justify the stringy approach to Wilson loops using various limits and
expansions: large size, large number of colors \cite{Hooft-1974,Veneziano-1976,MM-1981,Migdal-1984}, large number of space-time dimensions
\cite{Alvarez,Arvis-1983,Ambjorn-Makeenko-2016,Makeenko-2012-a}, Regge limit
\cite{Makeenko-2012-a,Makeenko-2009,Makeenko-2012-b}. Here we are interested in EST understood as an
effective theory describing the large-size limit. The first steps in this
direction were made by M.~L\"{u}scher, G.~M\"{u}nster, K.~Symanzik and
P.~Weisz \cite{Luscher:1980fr,Luscher:1980ac,Luscher:1980iy}. In the computation of
the first terms of the large-size expansion, one can approximate
$\mathcal{S}_{\mathrm{EST}}\left[  \Sigma\right]  $ by Nambu action
\begin{equation}
\mathcal{S}_{\text{Nambu}}\left(  \Sigma\right)  =\sigma_{0}S\left(
\Sigma\right)  \label{S-Nambu}
\end{equation}
where $S\left(  \Sigma\right)  $ is the area of surface $\Sigma$ and
$\sigma_{0}$ is the bare string tension (different from the renormalized
physical string tension $\sigma$ appearing in area law (\ref{Wilson-area-law}
)). If one wants to use EST for the calculation of higher-order terms of the
large-size expansion then $S_{\mathrm{EST}}\left[  \Sigma\right]  $ must be
understood as an infinite series containing all possible terms compatible with
the symmetries of the problem.

The theoretical work in EST has been proceeding in various directions including

\noindent$\bullet$ derivation of general constraints on terms appearing in EST action
$S_{\mathrm{EST}}\left[  \Sigma\right] $ 
\cite{Luscher:2004ib,Meyer:2006qx,Aharony:2009gg,Aharony:2010cx,Aharony:2010db,Aharony:2011gb,AFK-2012,AK-2013,BCGMP-12,GM-12},

\noindent$\bullet$ computation of loop corrections in EST for rectangular Wilson loops
\cite{BCGMP-12,Filk-preprint,Dietz-83,BCVG-2010,Billo:2011fd} and for other
closely related quantities like correlation functions of Polyakov lines and
spectra of closed and open strings \cite{Luscher:2004ib,Aharony:2009gg,Aharony:2010cx,Aharony:2010db,Aharony:2011gb,AFK-2012,AK-2013},

\noindent$\bullet$ analysis of string finite-width effects \cite{Luscher:1980iy,Gliozzi:2010zt,Meyer-2010}.

The predictions of EST have been successfully verified by many lattice Monte
Carlo tests (see \cite{BCGMP-12,BM-16,Athenodorou:2010cs,Athenodorou:2011rx,AT-2013,AT-2016-a,AT-2016-b,Pobylitsa-2016} and
references therein).

In the case when large-size limit $\left|  C\right|  \rightarrow\infty$ is
implemented as a uniform rescaling of contour $C$ with a large factor $\lambda$,
EST predicts the following structure of the large-size expansion for
polygonal contours $C$:

\begin{equation}
\ln W\left(  \lambda C\right)  \overset{\lambda\rightarrow\infty}{=}f_{\ln
}\left(  C\right)  \ln\lambda+\sum_{k\geq-2}\lambda^{-k}f_{k}\left(  C\right)
. \label{W-lambda-C-general-expansion-0}
\end{equation}
Here $f_{k}\left(  C\right)  $ ($k=\ln,-2,-1,0,\ldots$) are functionals of
contour $C$. EST provides much information on $f_{k}\left(  C\right)  $ but
some parameters controlling $f_{k}\left(  C\right)$ depend on the underlying
microscopic gauge theory (MGT).

Two leading terms of expansion (\ref{W-lambda-C-general-expansion-0}) are
associated with area $S\left(  C\right)  $ and
perimeter $L\left(  C\right)  $:
\begin{equation}
f_{-2}\left(  C\right)  =-\sigma S\left(  C\right)  , \label{f-minus-2-S}
\end{equation}
\begin{equation}
f_{-1}\left(  C\right)  =\rho L\left(  C\right)  . \label{f-minus-1-L}
\end{equation}
Here $\sigma$ is the string tension appearing in eq.~(\ref{Wilson-area-law}).
Parameters $\sigma,\rho$ are determined by MGT and cannot be computed in EST.
Parameter $\rho$ depends on the renormalization scheme used for Wilson loops
in MGT whereas string tension $\sigma$ is renormalization invariant in MGT.

Some properties of functionals $f_{k}\left(  C\right)  $ can be derived on
general grounds without a direct involvement of EST. For example, the
compatibility of expansion (\ref{W-lambda-C-general-expansion-0}) with
transitivity $W\left(  \left(  \lambda_{1}\lambda_{2}\right)  C\right)
=W\left(  \lambda_{1}\left(  \lambda_{2}C\right)  \right)  $ leads to relation
\begin{equation}
f_{\ln}\left(  C\right)  \ln\left(  \lambda_{1}\lambda_{2}\right)
+\sum_{k\geq-2}\left(  \lambda_{1}\lambda_{2}\right)  ^{-k}f_{k}\left(
C\right)  \,=f_{\ln}\left(  \lambda_{2}C\right)  \ln\lambda_{1}+\sum_{k\geq
-2}\lambda_{1}^{-k}f_{k}\left(  \lambda_{2}C\right)  \,
\end{equation}
which results in
\begin{equation}
f_{k}\left(  \lambda C\right)  =\lambda^{-k}f_{k}\left(  C\right)
\quad\left(  k\neq0\right)  , \label{f-n-scaling}
\end{equation}
\begin{equation}
f_{0}\left(  \lambda C\right)  =f_{0}\left(  C\right)  +f_{\ln}\left(
C\right)  \ln\lambda, \label{f-0-scaling}
\end{equation}
\begin{equation}
f_{\ln}\left(  \lambda C\right)  =f_{\ln}\left(  C\right)  .
\label{f-ln-scaling}
\end{equation}
Using EST, one can show that
\begin{equation}
f_{1}\left(  C\right)  =0\, \label{f-1-is-zero}
\end{equation}
(see section~\ref{Loop-expansion-section}).

\subsection{Simple example}

The primary subject of this paper is functional $f_{2}\left(  C\right)  $. In
the case of rectangular contours, $f_{2}\left(  C\right)  $ has been already
computed (see section~\ref{known-results-f2-section}). Our aim is to provide a
general computational scheme (including regularization and renormalization)
for the case of $f_{2}\left(  C\right)  $ with arbitrary polygons $C$ and to
perform an explicit calculation for the simplest case of triangles $C$.

Although term $\lambda^{-2}f_{2}\left(  C\right)  $ is strongly suppressed in
expansion (\ref{W-lambda-C-general-expansion-0}), this term plays a crucial
role in lattice tests of EST. The computation of this correction for general
polygons is also important for understanding the structure of renormalization
in EST (in the sense of renormalization in effective theories).

As for more dominant terms of expansion (\ref{W-lambda-C-general-expansion-0})
associated with $f_{\ln}\left(  C\right)  $ and $f_{0}\left(  C\right)  $,
these terms are well-known. A brief review of their properties can be found in
appendices~\ref{one-loop-EST-appendix},
\ref{Computation-Laplace-determinants-appendix}.

In this introductory section we would like to concentrate on functional
$f_{2}\left(  C\right)  $ and to minimize the involvement of terms $f_{\ln
}\left(  C\right)  $ and $f_{0}\left(  C\right)  $. Therefore it makes sense
to start from an expression combining several Wilson loops in such a way that
the role of term $f_{2}\left(  C\right)  \,$\ is enhanced whereas some of
dominant terms $f_{k}\left(  C\right)  $ cancel. An instructive example is
ratio
\begin{equation}
\frac{W\left(  \lambda\nu_{1}C\right)  W\left(  \lambda\nu_{2}C\right)
}{\left[  W\left(  \lambda\frac{\nu_{1}+\nu_{2}}{2}C\right)  \right]  ^{2}}
\label{W-ratio-1}
\end{equation}
taken in the limit $\lambda\rightarrow\infty$ at fixed $\nu_{1},\nu_{2}$.
Using properties (\ref{f-n-scaling}) -- (\ref{f-1-is-zero}) of expansion
(\ref{W-lambda-C-general-expansion-0}), we find
\begin{align}
&  \ln\frac{W\left(  \lambda\nu_{1}C\right)  W\left(  \lambda\nu_{2}C\right)
}{\left[  W\left(  \lambda\frac{\nu_{1}+\nu_{2}}{2}C\right)  \right]  ^{2}
}\overset{\lambda\rightarrow\infty}{=}-\frac{1}{2}\lambda^{2}\left(  \nu
_{1}-\nu_{2}\right)  ^{2}\sigma S\left(  C\right) \nonumber\\
&  +f_{\ln}\left(  C\right)  \ln\frac{4\nu_{1}\nu_{2}}{\left(  \nu_{1}+\nu
_{2}\right)  ^{2}}\nonumber\\
&  +\frac{1}{\lambda^{2}}\left[  \frac{1}{\nu_{1}^{2}}+\frac{1}{\nu_{2}^{2}
}-\frac{8}{\left(  \nu_{1}+\nu_{2}\right)  ^{2}}\right]  f_{2}\left(
C\right)  +O\left(  \lambda^{-3}\right)  . \label{W-ratio-2}
\end{align}
The LHS is a renormalization invariant combination of Wilson loops (which can
be computed, e.g. by lattice Monte Carlo simulations). On the RHS we have an
asymptotic series in inverse powers of $\lambda\rightarrow\infty$. Note that
terms $f_{-1}\left(  C\right)  $ and $f_{0}\left(  C\right)  $ cancel on the
RHS\ of (\ref{W-ratio-2}). The explicit expression for $f_{\ln}\left(
C\right)  $ is given by eq.~(\ref{f-ln}).

Expansion (\ref{W-ratio-2}) provides a good illustration of the role of the
$\lambda^{-2}f_{2}\left(  C\right)  $ correction. On the other hand, one can
construct more complicated renormalization invariant ratios of Wilson loops
whose large-size expansions involve also $f_{0}\left(  C\right) $: see
appendices~\ref{one-loop-EST-appendix},
\ref{Computation-Laplace-determinants-appendix}.

\subsection{Known results for $f_{2}\left(  C\right)  $}

\label{known-results-f2-section}

Dimensional arguments and scaling property (\ref{f-n-scaling}) lead to the
following structure of $f_{2}\left(  C\right)  $:
\begin{equation}
f_{2}\left(  C\right)  =\frac{1}{\sigma S\left(  C\right)  }g_{2}\left(
C\right)  \label{f2-via-g2}
\end{equation}
where $\sigma$ is the string tension appearing in area law
(\ref{Wilson-area-law}) and $g_{2}\left(  C\right)  $ is a dimensionless
function depending only on the geometry of polygon $C$.

For rectangular contours, $f_{2}\left(  C\right)  $ was computed
in refs.~\cite{Filk-preprint,Dietz-83} with an unfortunate arithmetic error which was later
detected and corrected in refs.~\cite{BCVG-2010,Billo:2011fd}. For
rectangle $C_{\text{rectangle}}\left(  L_{1},L_{2}\right)  $ with
sides\ $L_{1},L_{2}$ one has
\begin{align}
g_{2}\left(  C_{\text{rectangle}}\left(  L_{1},L_{2}\right)  \right)   &
=\left(  \frac{\pi}{24}\right)  ^{2}\left(  D-2\right)\left[  2  \left(
\frac{L_{1}}{L_{2}}\right)  ^{2}E_{4}\left(  i\frac{L_{1}}{L_{2}}\right)
\right. \nonumber\\
&  \left.  -\frac{ D-6  }{2}E_{2}\left(
i\frac{L_{1}}{L_{2}}\right)  E_{2}\left(  i\frac{L_{2}}{L_{1}}\right)
\right]  \label{g-2-rect-res}
\end{align}
where $D$ is the space-time dimension and $E_{n}\left(  z\right)  $ are
Eisenstein series.

In this work we compute $f_{2}\left(  C\right)  $ for the case of triangular
loops $C$. For triangle $C_{\text{triangle}}\left(  \theta_{1},\theta
_{2},\theta_{3};S\right)  $ with interior angles $\theta_{1},\theta_{2}
,\theta_{3}$ and with area $S$ we find (see section~\ref{final-result-section})
\begin{equation}
g_{2}\left(  C_{\text{triangle}}\left(  \theta_{1},\theta_{2},\theta
_{3};S\right)  \right)  =\frac{1}{8}Y\left(  1-Y\right)
\end{equation}
where
\begin{equation}
Y=\frac{D-2}{24}\prod_{k=1}^{3}\left(  \frac{\pi}{\theta_{k}}-1\right)  .
\end{equation}
According to eq.~(\ref{f2-via-g2}), the full explicit expression for
$f_{2}\left(  C_{\text{triangle}}\right)  $ is
\begin{gather}
f_{2}\left(  C_{\text{triangle}}\left(  \theta_{1},\theta_{2}
,\theta_{3};S\right)  \right)  =\frac{1}{\sigma S}\frac{D-2}
{192}\nonumber\\
\times\left[  \prod_{k=1}^{3}\left(  \frac{\pi}{\theta_{k}}-1\right)  \right]
\left[  1-\frac{D-2}{24}\prod_{k=1}^{3}\left(  \frac{\pi}{\theta_{k}
}-1\right)  \right]  . \label{f2-triangle-re-intro}
\end{gather}
\qquad\qquad\qquad

\subsection{Outline of the work}

The main part of this work concentrates on the computation of $f_{2}\left(
C\right)  $ in large-size expansion (\ref{W-lambda-C-general-expansion-0}).
Working with ratios (\ref{W-ratio-2}) of Wilson loops, one can eliminate more
dominant term $f_{0}\left(  C\right)  $ from the consideration. Nevertheless
for completeness a review of well-known results for terms $f_{\ln}\left(
C\right)  $ and $f_{0}\left(  C\right)  $ is provided
in appendices~ \ref{Laplace-determinant-properties-appendix}, \ref{one-loop-EST-appendix},
\ref{Computation-Laplace-determinants-appendix}.

In section~\ref{section-Wilson-loops-EST} we sketch basic formulas of the loop
expansion in EST. This part of the work is common for our triangular case and
for the rectangular case studied in refs.~\cite{Filk-preprint,Dietz-83}.
Correction $f_{2}\left(  C\right)  $ is given by a
figure-eight-like Feynman diagram for a two-dimensional field theory of
$\left(  \partial\phi\right)  ^{2}+\left(  \partial\phi\right)  ^{4}$ type
(see eq.~(\ref{S-Nambu-expansion-tr-gauge})) in the region bounded by contour
$C$ with zero boundary conditions. We use notation $\left\langle
\mathcal{S}_{4}\right\rangle $ (see eqs.~(\ref{S-4-def-0}),
(\ref{S-4-average-0})) for the integral associated with this Feynman diagram.
In principle, $\left\langle \mathcal{S}_{4}\right\rangle $ represents the
result for $f_{2}\left(  C\right)  $ (up to the sign). However, this Feynman
diagram has ultraviolet divergences.

The deviation of the current work from refs.~\cite{Filk-preprint,Dietz-83}
starts when it comes to the regularization of ultraviolet
divergences in $\left\langle \mathcal{S}_{4}\right\rangle $. Unfortunately the
regularization used in refs.~\cite{Filk-preprint,Dietz-83} can be
applied only to the case of rectangular contours $C$. In our work we suggest
another regularization which can be used for arbitrary polygons.
In section~\ref{renormalization-introductory-section} we describe the difference between
the regularization used in refs.~\cite{Filk-preprint,Dietz-83} and our
regularization. Our renormalization procedure operationally consists of two
steps:
\begin{equation}
\left\langle \mathcal{S}_{4}\right\rangle \overset{\text{step 1}
}{\longrightarrow}\left\langle \mathcal{S}_{4}\right\rangle ^{\text{ren,1}
}\overset{\text{step 2}}{\longrightarrow}\left\langle \mathcal{S}
_{4}\right\rangle ^{\text{ren}}\equiv-f_{2}\left(  C\right)  .
\label{renormalization-chain}
\end{equation}
Here $\left\langle \mathcal{S}_{4}\right\rangle ^{\text{ren}}$ is completely
free of ultraviolet divergences and provides the result for $f_{2}\left(
C\right)  $ (see eq.~(\ref{f2-C-S-ren})) whereas the intermediate quantity
$\left\langle \mathcal{S}_{4}\right\rangle ^{\text{ren,1}}$ (introduced in eq.~(\ref{S4-average-ren-1-real})) is only partly
renormalized and still has ultraviolet divergences.

Our calculation is based on Schwarz-Christoffel (SC)\ mapping for polygon $C$
and on other powerful methods of complex analysis.
In section~\ref{complex-representation-section} we introduce a complexified
representation for $\left\langle \mathcal{S}_{4}\right\rangle $ and use it in
order to perform the first step of renormalization $\left\langle
\mathcal{S}_{4}\right\rangle \rightarrow\left\langle \mathcal{S}
_{4}\right\rangle ^{\text{ren,1}}$ (with technical details placed
 in appendix~\ref{Appendix-Diagonal-limit}).

In section~\ref{SC-mapping-section} the derived expression for
$\left\langle \mathcal{S}_{4}\right\rangle ^{\text{ren,1}}$ is rewritten in terms of SC\ mapping.

In section~\ref{Analytical-regularization-section} we describe the second step of
renormalization procedure $\left\langle \mathcal{S}_{4}\right\rangle
^{\text{ren,1}}\rightarrow\left\langle S\right\rangle ^{\text{ren}}$ based on
the analytical regularization of the SC\ representation for $\left\langle
\mathcal{S}_{4}\right\rangle ^{\text{ren,1}}$, assuming the general case of
arbitrary polygons $C$.

In section~\ref{case-of-triangles-section} we apply the general formula derived
for $\left\langle S\right\rangle ^{\text{ren}}$ with an arbitrary polygon to
the case of triangles. We show that in the triangular case all integrals
involved in the computation of $\left\langle S\right\rangle ^{\text{ren}}$ can
be computed explicitly and derive final expression (\ref{f2-triangle-re-intro})
for $f_{2}\left(  C\right)  =-\left\langle S\right\rangle ^{\text{ren}}$.
The technical details of the computation are placed
in appendices~\ref{Appendix-SC-mapping-triangles}, \ref{Appendix-K-1-2-calculation},
\ref{Appendix-K-P-2-calculation}. For completeness we also provide information
about $f_{0}\left(  C\right)  $ with triangular contours $C$
in appendix~\ref{triangle-appendix}.

\section{Wilson loops in EST}

\setcounter{equation}{0} 

\label{section-Wilson-loops-EST}

\subsection{Expansion of Nambu action}

The first steps of our work are identical to those
of refs.~\cite{Filk-preprint,Dietz-83}. For the computation of the first orders
of the large-size expansion of Wilson loops (including $f_{2}\left(  C\right)
$), one can replace full EST\ action $\mathcal{S}_{\mathrm{EST}}\left[
\Sigma\right]$   in eq.~(\ref{W-C-EST}) by Nambu action (non-Nambu terms of effective action
$\mathcal{S}_{\mathrm{EST}}\left[  \Sigma\right]  $ become essential starting
from the computation of correction $f_{3}\left(  C\right)  $ generated by the
boundary action \cite{Luscher:2004ib,Aharony:2010cx,BCGMP-12}).
Thus for our aim we can approximate
\begin{equation}
W\left(  C\right)  \rightarrow\mathrm{const}\int_{\partial\Sigma=C}D\Sigma
\exp\left[  -\mathcal{S}_{\text{Nambu}}\left(  \Sigma\right)  \right]
\label{W-int-sigma}
\end{equation}
with $\mathcal{S}_{\text{Nambu}}\left(  \Sigma\right)  $ given by eq.
(\ref{S-Nambu}). Assuming parametrization $X^{a}\left(  x^{1},x^{2}\right)  $
of surface $\Sigma$, one has
\begin{equation}
\mathcal{S}_{\text{Nambu}}\left(  \Sigma\right)  =\sigma_{0}\int dx^{1}
dx^{2}\sqrt{\det_{\mu\nu}\left(  Q_{\mu\nu}\right)  },
\end{equation}
\begin{equation}
Q_{\mu\nu}=\sum_{a=1}^{D}\frac{\partial X^{a}}{\partial x^{\mu}}\frac{\partial
X^{a}}{\partial x^{\nu}}.
\end{equation}
In EST\ (like in the theory of fundamental boson strings) the reparametrization
invariance must be treated as a gauge symmetry. In the case of flat Wilson
contours $C$ lying in the $X^{1},X^{2}$ plane, it is convenient to use the
planar gauge (sometimes called static gauge):
\begin{equation}
X^{a}\left(  x^{1},x^{2}\right)  =x^{a}\quad\left(  a=1,2\right)  ,
\end{equation}
\begin{equation}
X^{a}\left(  x^{1},x^{2}\right)  =Y^{a}\left(  x^{1},x^{2}\right)
\quad\left(  a=3,4,\ldots D\right)  .
\end{equation}

Expanding up to quartic terms, we find in this gauge
\begin{align}
\sqrt{\det_{\mu\nu}\left(  Q_{\mu\nu}\right)  }\,  &  =1+\frac{1}{2}
\sum\limits_{\mu=1}^{2}\sum\limits_{a=3}^{D}\left(  \frac{\partial Y^{a}
}{\partial x^{\mu}}\right)  ^{2}\nonumber\\
&  +\sum\limits_{\mu,\nu=1}^{2}\sum\limits_{a,b=3}^{D}\left[  \frac{1}
{8}\left(  \frac{\partial Y^{a}}{\partial x^{\mu}}\right)  ^{2}\left(
\frac{\partial Y^{b}}{\partial x^{\nu}}\right)  ^{2}-\frac{1}{4}\frac{\partial
Y^{a}}{\partial x^{\mu}}\frac{\partial Y^{b}}{\partial x^{\mu}}\frac{\partial
Y^{a}}{\partial x^{\nu}}\frac{\partial Y^{b}}{\partial x^{\nu}}\right]
+\ldots
\end{align}
We use notation $U\left(  C\right)  $ for the flat region in $x^{1},x^{2}$
plane bounded by Wilson contour $C$.

Thus
\begin{gather}
\frac{1}{\sigma_{0}}\mathcal{S}_{\text{Nambu}}=\int_{U\left(  C\right)  }
d^{2}x\sqrt{\det_{\mu\nu}\left(  Q_{\mu\nu}\right)  }\approx S\left(
C\right)  +\frac{1}{2}\int_{U\left(  C\right)  }d^{2}x\sum\limits_{\mu=1}
^{2}\sum\limits_{a=3}^{D}\left(  \frac{\partial Y^{a}}{\partial x^{\mu}
}\right)  ^{2}\nonumber\\
+\int_{U\left(  C\right)  }d^{2}x\sum\limits_{\mu,\nu=1}^{2}\sum
\limits_{a,b=3}^{D}\left[  \frac{1}{8}\left(  \frac{\partial Y^{a}}{\partial
x^{\mu}}\right)  ^{2}\left(  \frac{\partial Y^{b}}{\partial x^{\nu}}\right)
^{2}-\frac{1}{4}\frac{\partial Y^{a}}{\partial x^{\mu}}\frac{\partial Y^{b}
}{\partial x^{\mu}}\frac{\partial Y^{a}}{\partial x^{\nu}}\frac{\partial
Y^{b}}{\partial x^{\nu}}\right]  +\ldots\label{S-Nambu-expansion-tr-gauge}
\end{gather}
Here $S\left(  C\right)  $ is the area of region $U\left(  C\right)  $ bounded
by flat contour $C$. One can rewrite this expansion in the form
\begin{equation}
\mathcal{S}_{\text{Nambu}}=\mathcal{S}_{0}+\mathcal{S}_{2}+\mathcal{S}
_{4}+\ldots\label{S-S2-S4}
\end{equation}
where
\begin{equation}
\mathcal{S}_{0}=\sigma_{0}S\left(  C\right)  ,
\end{equation}
\begin{equation}
\mathcal{S}_{2}=\frac{1}{2}\sigma_{0}\int_{U\left(  C\right)  }d^{2}
x\sum\limits_{\mu=1}^{2}\sum\limits_{a=3}^{D}\left(  \frac{\partial Y^{a}
}{\partial x^{\mu}}\right)  ^{2},
\end{equation}
\begin{equation}
\mathcal{S}_{4}=\sigma_{0}\int_{U\left(  C\right)  }d^{2}x\sum\limits_{\mu
,\nu=1}^{2}\sum\limits_{a,b=3}^{D}\left[  \frac{1}{8}\left(  \frac{\partial
Y^{a}}{\partial x^{\mu}}\right)  ^{2}\left(  \frac{\partial Y^{b}}{\partial
x^{\nu}}\right)  ^{2}-\frac{1}{4}\frac{\partial Y^{a}}{\partial x^{\mu}
}\frac{\partial Y^{b}}{\partial x^{\mu}}\frac{\partial Y^{a}}{\partial x^{\nu
}}\frac{\partial Y^{b}}{\partial x^{\nu}}\right]  . \label{S-4-def-0}
\end{equation}

\subsection{Loop expansion in EST}

\label{Loop-expansion-section}

The large-size expansion for Wilson loops is constructed by applying
the steepest-descent method to functional integral (\ref{W-int-sigma}). Expanding
in small fluctuations of $\Sigma$ near the minimal surface spanned on contour
$C$, one finds in the 1-loop approximation \cite{Luscher:1980fr,Luscher:1980ac,Luscher:1980iy}:
\begin{align}
\int_{\partial\Sigma=C}D\Sigma\exp\left[  -\mathcal{S}_{\text{Nambu}}\left(
\Sigma\right)  \right]   &  =\mathrm{const}\,\left[  \mathrm{Det}
_{\mathrm{reg}}\left[  -\Delta\left(  C\right)  \right]  \right]  ^{-\left(
D-2\right)  /2}\exp\left[  -\sigma_{0}S\left(  C\right)  \right]
\,\nonumber\\
&  \times\left[  1+O\left(  \frac{1}{\sigma_{0}}\right)  \right]  .
\label{int-sigma-one-loop}
\end{align}
Here $\mathrm{Det}_{\mathrm{reg}}\left[  -\Delta\left(  C\right)  \right]  $
is the functional determinant of Laplace operator (referred as
\emph{Laplace determinant} for brevity) with zero (Dirichlet) boundary
conditions on contour $C$ and with properly regularized ultraviolet divergences.
In this oversimplified approach, at intermediate stages of the calculation we
formally deal with the loop expansion in small parameter $1/\sigma
_{0}\rightarrow0$ at fixed contour $C$ but in the end this formal
$1/\sigma_{0}$ expansion can be rearranged as a large-size expansion for
Wilson loops at fixed physical (renormalized) string tension $\sigma$. The
real situation is more complicated because

\noindent$\bullet$ loop corrections of EST have ultraviolet divergences which must be
renormalized (in the sense of renormalization in effective theories),

\noindent$\bullet$ microscopic gauge theory (MGT) has its own ultraviolet\ divergences and renormalization.

Fortunately, these complications play a limited role if one is interested in
the EST computation of the first terms of  large-size expansion
(\ref{W-lambda-C-general-expansion-0}).

For the computation of $f_{2}\left(  C\right)  $ one needs only the Nambu term
of the full effective action. The loop expansion of EST with Nambu action\ naively
goes in even powers of $1/\sigma_{0}$ which results in even terms
$\lambda^{-2n}f_{2n}\left(  C\right)  $ of large-size expansion
(\ref{W-lambda-C-general-expansion-0}). This naive argument explains the
vanishing of odd term $f_{1}\left(  C\right)  $ (\ref{f-1-is-zero}) but a
serious proof of (\ref{f-1-is-zero}) requires more effort since divergences
and their renormalization in EST may generate odd and non-analytical terms
like $f_{-1}\left(  C\right)  $ and $f_{\ln}\left(  C\right)  $.

Anyway the concept of the naive and formal loop expansion in small parameter
$1/\sigma_{0}$ may be helpful (at least operationally) for the computation of
$f_{k}\left(  C\right)  $ in EST and within this framework one has the
following correspondence between loop counting and $f_{k}\left(  C\right)  $
computation:
\begin{align}
\text{1 loop }  &  \Longrightarrow\,f_{\ln}\left(  C\right)  \ln\lambda
+f_{0}\left(  C\right)  ,\label{one-loop-terms}\\
\text{2 loops }  &  \Longrightarrow\,\lambda^{-2}f_{2}\left(  C\right)  .
\label{two-loop-terms}
\end{align}

The main subject of this paper is 2-loop correction $f_{2}\left(  C\right)
$. Technically (but not conceptually) the computation of $f_{2}\left(
C\right)  $ can be well separated from the computation of 1-loop terms
$f_{\ln}\left(  C\right)  \ln\lambda+f_{0}\left(  C\right)  $. Therefore in
subsequent sections we concentrate on $f_{2}\left(  C\right)  $ whereas a
detailed discussion of 1-loop terms is placed
to appendices~\ref{one-loop-EST-appendix}, \ref{Computation-Laplace-determinants-appendix}.

In 1-loop approximation (\ref{int-sigma-one-loop}) there appears Laplace
determinant which has area, perimeter and cusp divergences
(see appendices~\ref{heat-kernel-appendix}, \ref{Laplace-determinant-renormalization-appendix}).
Therefore the completion of the 1-loop computation (i.e. the computation of
$f_{\ln}\left(  C\right)  $, $f_{0}\left(  C\right)  $) includes two parts:

\noindent 1) renormalization of Laplace determinant
(appendix~\ref{Laplace-determinant-renormalization-appendix}) using the heat-kernel
expansion (appendix~\ref{heat-kernel-appendix}),

\noindent 2) computation of the renormalized Laplace determinant for arbitrary polygons
(appendix~\ref{Computation-Laplace-determinants-appendix}) using
Schwarz-Christoffel (SC)\ mapping.

Both parts of this work are well described in literature so that we simply combine all
known pieces together
in appendices~\ref{Laplace-determinant-properties-appendix}, \ref{one-loop-EST-appendix},
\ref{Computation-Laplace-determinants-appendix}.

An explicit expression for $f_{\ln}\left(  C\right)  $ can be obtained already
at the first step (renormalization of Laplace determinants) --- see eq.
(\ref{f-ln}). As for $f_{0}\left(  C\right)  $, the situation is more subtle
because single Wilson loops $W\left(  C\right)  $ and corresponding
functionals $f_{0}\left(  C\right)  $ are not renormalization invariant
quantities in MGT. EST provides unambiguous expressions only for
renormalization invariant combinations
\begin{equation}
\sum_{a=1}^{N}m_{a}f_{0}\left(  C_{a}\right)  \label{m-f-C-combination}
\end{equation}
where contours $C_{a}$ and coefficients $m_{a}$ must obey certain balance
conditions (\ref{balance-vertices-0}), (\ref{balance-perimeter-0}),
(\ref{balance-angles-0}). Once these balance conditions are satisfied, linear
combination (\ref{m-f-C-combination}) is given by eq.~(\ref{m-f0-C-sum-res})
derived in appendix~\ref{Computation-Laplace-determinants-appendix} in terms
of parameters of SC\ mapping.

In the special case when all polygons $C_{a}$ in linear combination
(\ref{m-f-C-combination}) are triangles one can derive a simpler expression
for (\ref{m-f-C-combination}) directly in terms of geometric parameters of
triangles --- see eq.~(\ref{m-f0-C-sum-triangles-res}).

\subsection{2-loop EST correction}

Using action (\ref{S-S2-S4}), we compute the functional integral of EST in
the 2-loop approximation:

\begin{align}
W\left(  C\right)   &  \rightarrow\int DY\exp\left\{  -\left[  \mathcal{S}
_{0}+\mathcal{S}_{2}\left(  Y\right)  +\mathcal{S}_{4}\left(  Y\right)
\right]  \right\} \nonumber\\
&  \rightarrow\exp\left(  -\mathcal{S}_{0}\right)  \int DY\left[
1-\mathcal{S}_{4}\left(  Y\right)  \right]  \exp\left[  -\mathcal{S}
_{2}\left(  Y\right)  \right] \nonumber\\
&  =\exp\left(  -\mathcal{S}_{0}\right)  \left(  1-\left\langle \mathcal{S}
_{4}\right\rangle \right)  \int DY\exp\left[  -\mathcal{S}_{2}\left(
Y\right)  \right]  .
\end{align}
Here
\begin{equation}
\left\langle \mathcal{S}_{4}\right\rangle =\frac{\int DY\,\mathcal{S}
_{4}\left(  Y\right)  \exp\left[  -\mathcal{S}_{2}\left(  Y\right)  \right]
}{\int DY\,\exp\left[  -\mathcal{S}_{2}\left(  Y\right)  \right]  }.
\label{S-4-average-0}
\end{equation}
Thus in the 2-loop approximation
\begin{equation}
\left\langle W\left(  C\right)  \right\rangle ^{\text{2-loop}}=\exp\left(
-\mathcal{S}_{0}\right)  \left[  1-\left\langle \mathcal{S}_{4}\right\rangle
\right]  \int DY\exp\left[  -\mathcal{S}_{2}\left(  Y\right)  \right]  .
\end{equation}
Comparing this with the 1-loop approximation
\begin{equation}
\left\langle W\left(  C\right)  \right\rangle ^{\text{1-loop}}=\exp\left(
-\mathcal{S}_{0}\right)  \int DY\exp\left[  -\mathcal{S}_{2}\left(  Y\right)
\right]  , \label{W-one-loop-Gaussian}
\end{equation}
we see that
\begin{equation}
\left\langle W\left(  C\right)  \right\rangle ^{\text{2-loop}}=\left\langle
W\left(  C\right)  \right\rangle ^{\text{1-loop}}\left(  1-\left\langle
\mathcal{S}_{4}\right\rangle \right)  , \label{W-correction-via-S4}
\end{equation}
which leads to
\begin{equation}
\ln\left\langle W\left(  C\right)  \right\rangle ^{\text{2-loop}}
\rightarrow\ln\left\langle W\left(  C\right)  \right\rangle ^{\text{1-loop}
}-\left\langle \mathcal{S}_{4}\right\rangle .
\end{equation}
In fact, one should use the renormalized version of this formula:
\begin{equation}
\ln\left\langle W\left(  C\right)  \right\rangle _{\text{ren}}^{\text{2-loop}
}=\ln\left\langle W\left(  C\right)  \right\rangle _{\text{ren}}
^{\text{1-loop}}-\left\langle \mathcal{S}_{4}\right\rangle ^{\text{ren}}.
\end{equation}

Comparing this with expansion (\ref{W-lambda-C-general-expansion-0}), we see
that
\begin{equation}
\lambda^{-2}f_{2}\left(  C\right)  =-\left\langle \mathcal{S}_{4}\right\rangle
_{\lambda C}^{\text{ren}}.
\end{equation}
Taking into account eq.~(\ref{f-n-scaling}), we obtain
\begin{equation}
f_{2}\left(  C\right)  =-\left\langle \mathcal{S}_{4}\right\rangle
_{C}^{\text{ren}}. \label{f2-C-S-ren}
\end{equation}
This relation is used in the final step of our computational program (\ref{renormalization-chain}).

Using expression (\ref{S-4-def-0}) for $\mathcal{S}_{4}$, we compute Gaussian
integral (\ref{S-4-average-0})
\begin{align}
\frac{1}{\sigma_{0}}\left\langle \mathcal{S}_{4}\right\rangle  &  =\frac{1}
{8}\int_{U\left(  C\right)  }d^{2}x\sum\limits_{\mu,\nu=1}^{2}\sum
\limits_{a,b=3}^{D}\left\langle \frac{\partial Y^{a}}{\partial x^{\mu}
}\frac{\partial Y^{a}}{\partial x^{\mu}}\frac{\partial Y^{b}}{\partial x^{\nu
}}\frac{\partial Y^{b}}{\partial x^{\nu}}\right\rangle \nonumber\\
&  -\frac{1}{4}\int_{U\left(  C\right)  }d^{2}x\sum\limits_{\mu,\nu=1}^{2}
\sum\limits_{a,b=3}^{D}\left\langle \frac{\partial Y^{a}}{\partial x^{\mu}
}\frac{\partial Y^{b}}{\partial x^{\mu}}\frac{\partial Y^{a}}{\partial x^{\nu
}}\frac{\partial Y^{b}}{\partial x^{\nu}}\right\rangle
.\label{average-S4-calc-1}
\end{align}
Now Wick theorem yields a figure-eight-like Feynman diagram
\begin{align}
\frac{1}{\sigma_{0}}\left\langle \mathcal{S}_{4}\right\rangle  &  =\frac{1}
{8}\int_{U\left(  C\right)  }d^{2}x\sum\limits_{\mu,\nu=1}^{2}\sum
\limits_{a,b=3}^{D}\left[  \left\langle \frac{\partial Y^{a}}{\partial x^{\mu
}}\frac{\partial Y^{a}}{\partial x^{\mu}}\right\rangle \left\langle
\frac{\partial Y^{b}}{\partial x^{\nu}}\frac{\partial Y^{b}}{\partial x^{\nu}
}\right\rangle \right.  \\
&  \left.  +2\left\langle \frac{\partial Y^{a}}{\partial x^{\mu}
}\frac{\partial Y^{b}}{\partial x^{\nu}}\right\rangle \left\langle
\frac{\partial Y^{a}}{\partial x^{\mu}}\frac{\partial Y^{b}}{\partial x^{\nu}
}\right\rangle \right]  \nonumber\\
&  -\frac{1}{4}\int_{U\left(  C\right)  }d^{2}x\sum\limits_{\mu,\nu=1}^{2}
\sum\limits_{a,b=3}^{D}\left[  \left\langle \frac{\partial Y^{a}}{\partial
x^{\mu}}\frac{\partial Y^{b}}{\partial x^{\mu}}\right\rangle \left\langle
\frac{\partial Y^{a}}{\partial x^{\nu}}\frac{\partial Y^{b}}{\partial x^{\nu}
}\right\rangle \right.  \nonumber\\
&  \left.  +\left\langle \frac{\partial Y^{a}}{\partial x^{\mu}}\frac{\partial
Y^{a}}{\partial x^{\nu}}\right\rangle \left\langle \frac{\partial Y^{b}
}{\partial x^{\mu}}\frac{\partial Y^{b}}{\partial x^{\nu}}\right\rangle
+\left\langle \frac{\partial Y^{a}}{\partial x^{\mu}}\frac{\partial Y^{b}
}{\partial x^{\nu}}\right\rangle \left\langle \frac{\partial Y^{b}}{\partial
x^{\mu}}\frac{\partial Y^{a}}{\partial x^{\nu}}\right\rangle \right]
.\label{average-S4-calc-2}
\end{align}
with the propagator
\begin{equation}
\left\langle \frac{\partial Y^{a}}{\partial x^{\mu}}\frac{\partial Y^{b}
}{\partial x^{\nu}}\right\rangle =-\frac{1}{\sigma_{0}}\delta_{ab}G_{\mu\nu
}\label{average-dY-dY-via-G}
\end{equation}
where
\begin{equation}
G_{\mu\nu}\left(  x\right)  =-\left\langle x\right|  \partial_{\mu}
\frac{1}{\Delta\left(  C\right)  }\partial_{\nu}\left|  x\right\rangle
,\label{G-mu-nu-def}
\end{equation}
\begin{equation}
\left\langle y\right|  \partial_{\mu}\frac{1}{\Delta\left(  C\right)
}\partial_{\nu}\left|  x\right\rangle =-\frac{\partial}{\partial y^{\mu}
}\frac{\partial}{\partial x^{\nu}}\left\langle y\right|  \frac{1}
{\Delta\left(  C\right)  }\left|  x\right\rangle \label{y-x-ME}
\end{equation}
and
\begin{equation}
\left\langle y\right|  \frac{1}{\Delta\left(  C\right)  }\left|
x\right\rangle
\end{equation}
is Green function of Laplace operator with Dirichlet boundary condition:
\begin{equation}
\sum_{\mu=1}^{2}\frac{\partial^{2}}{\partial y^{\mu}\partial y^{\mu}
}\left\langle y\right|  \frac{1}{\Delta\left(  C\right)  }\left|
x\right\rangle =\delta\left(  y-x\right)  ,
\end{equation}
\begin{equation}
\left.  \left\langle y\right|  \frac{1}{\Delta\left(  C\right)  }\left|
x\right\rangle \right|  _{y\in C}=0,
\end{equation}
\begin{equation}
\left\langle y\right|  \frac{1}{\Delta\left(  C\right)  }\left|
x\right\rangle =\left\langle x\right|  \frac{1}{\Delta\left(  C\right)
}\left|  y\right\rangle .
\end{equation}

Now we insert eq.~(\ref{average-dY-dY-via-G}) into eq.
(\ref{average-S4-calc-2})
\begin{align}
\sigma_{0}\left\langle \mathcal{S}_{4}\right\rangle  &  =\frac{D-2}{8}
\int_{U\left(  C\right)  }d^{2}x\sum\limits_{\mu,\nu=1}^{2}\left[  \left(
D-2\right)  G_{\mu\mu}G_{\nu\nu}+2G_{\mu\nu}G_{\mu\nu}\right] \nonumber\\
&  -\frac{D-2}{4}\int_{U\left(  C\right)  }d^{2}x\sum\limits_{\mu,\nu=1}
^{2}\left[  G_{\mu\mu}G_{\nu\nu}+\left(  D-1\right)  G_{\mu\nu}G_{\mu\nu
}\right]  .
\end{align}
Thus
\begin{equation}
\left\langle \mathcal{S}_{4}\right\rangle =-\frac{D-2}{8\sigma_{0}}
\int_{U\left(  C\right)  }d^{2}x\sum\limits_{\mu,\nu=1}^{2}\left[  \left(
4-D\right)  G_{\mu\mu}G_{\nu\nu}+2\left(  D-2\right)  G_{\mu\nu}G_{\mu\nu
}\right]  . \label{S-4-average-2}
\end{equation}
One should keep in mind that the RHS is plagued by divergences. Indeed,
nondiagonal matrix element (\ref{y-x-ME}) has a singularity at $y\rightarrow
x$ so that quantity $G_{\mu\nu}\left(  x\right)  $ is divergent for any $x$.
This means that on the RHS\ of eq.~(\ref{S-4-average-2}) the integrand is
divergent at any point $x$. Nevertheless eq.~(\ref{S-4-average-2}) provides a
good starting point for the discussion of the renormalization.

\section{Renormalization}
\setcounter{equation}{0} 

\label{renormalization-introductory-section}

\subsection{Rectangular case: method of K. Dietz and T. Filk}

The naive non-renormalized expression for the 2-loop correction
(\ref{S-4-average-2}) allows for arbitrary polygons $C$ and in this sense is
common for the current work concentrating on triangles and
for refs.~\cite{Filk-preprint,Dietz-83} devoted to rectangles. But starting from
eq.~(\ref{S-4-average-2}) the paths of this work
and refs.~\cite{Filk-preprint,Dietz-83} diverge. The main reason is
that the authors of refs.~\cite{Filk-preprint,Dietz-83} use an ultraviolet\ regularization
specific for the rectangular case which cannot be generalized to other polygons.
The regularization of refs.~\cite{Filk-preprint,Dietz-83} is based on
the explicit diagonalization of Laplace operator acting in rectangle
\begin{equation}
0\leq x^{\mu}\leq L_{\mu}\quad\left(  \mu=1,2\right)
\end{equation}
with Dirichlet boundary conditions. The spectral problem
\begin{equation}
-\Delta\psi_{mn}=\lambda_{mn}\psi_{mn}
\end{equation}
has the obvious solution
\begin{equation}
\lambda_{mn}=\pi^{2}\left(  \frac{m^{2}}{L_{1}^{2}}+\frac{n^{2}}{L_{2}^{2}
}\right)  , \label{lambda-m-n}
\end{equation}
\begin{equation}
\psi_{mn}=\frac{2}{\sqrt{L_{1}L_{2}}}\sin\left(  \frac{\pi mx^{1}}{L_{1}
}\right)  \sin\left(  \frac{\pi nx^{2}}{L_{2}}\right)  . \label{psi-m-n}
\end{equation}
The method of refs.~\cite{Filk-preprint,Dietz-83} is based on

\noindent 1) formal spectral decomposition of $G_{\mu\nu}$ (\ref{G-mu-nu-def}) in terms
of $\lambda_{mn}$, $\psi_{mn}$ ignoring the divergence of $G_{\mu\nu}$,

\noindent 2) insertion of this formal expression for $G_{\mu\nu}$ into the RHS\ of eq.
(\ref{S-4-average-2}),

\noindent 3) formal integration over $x$ in eq.~(\ref{S-4-average-2}).

As a result, one obtains an expression for $\left\langle \mathcal{S}
_{4}\right\rangle $ in terms of divergent series
\begin{equation}
\sum_{m_{1}m_{2}\ldots m_{n}}\frac{R_{1}\left(  m_{1},m_{2},\ldots
,m_{n}\right)  }{R_{2}\left(  m_{1},m_{2},\ldots,m_{n}\right)  }
\label{Q-sum-DF}
\end{equation}
where $R_{a}\left(  m_{1},m_{2},\ldots,m_{n}\right)  $ are some polynomials of
integer $m_{k}$ (originating from parameters $m$, $n$ of eigenvalues
$\lambda_{mn}$). Formal series (\ref{Q-sum-DF}) are divergent.
Refs.~\cite{Filk-preprint,Dietz-83} use analytical regularization for
numerators $R_{1}$ so that the problem reduces to the computation of sums
\begin{equation}
\sum_{m_{1}m_{2}\ldots m_{n}}\frac{m_{1}^{\alpha_{1}}m_{2}^{\alpha_{2}}\ldots
m_{n}^{\alpha_{n}}}{R_{2}\left(  m_{1},m_{2},\ldots,m_{n}\right)
}\label{alpha-sums-DF}
\end{equation}
in convergence region of $\alpha_{k}$-space with a subsequent analytical
continuation in $\alpha_{k}$ to integer values corresponding to monomials of
the original numerator $R_{1}\left(  m_{1},m_{2},\ldots,m_{n}\right)  $ in eq.
(\ref{Q-sum-DF}).

This approach allows for a complete analytical
computation of $f_{2}\left(  C\right)  $ for rectangles $C$ with result
(\ref{g-2-rect-res}). However, the method is based on the explicit expressions
for spectrum (\ref{lambda-m-n}) and for eigenfunctions (\ref{psi-m-n}) and
therefore is limited to the case of rectangles.

\subsection{General polygons: Renormalization based on SC\ mapping}

As was discussed above, the analytical regularization
of refs.~\cite{Filk-preprint,Dietz-83} was designed for rectangles. Our aim is to construct a regularization (and renormalization)
procedure for $f_{2}\left(  C\right)  $ with arbitrary polygons $C$. We also want
this regularization method to be efficient for practical computations. The
first hint for the construction of this regularization procedure comes from
the lessons of 1-loop EST corrections. As discussed
in appendix~\ref{Computation-Laplace-determinants-appendix}, the explicit analytical
computation of 1-loop correction $f_{0}\left(  C\right)  $ is based on
SC\ mapping for polygon $C$. This observation suggests that we should try to
construct the regularization and the renormalization for the 2-loop correction $f_{2}\left(  C\right)  $
also in terms of SC\ mapping.

The next observation is that the naive divergent expression
(\ref{S-4-average-2}) has two types of divergences:

1) The integrand is divergent at any point $x$ because $G_{\mu\nu}\left(
x\right)  $ is divergent at any $x.$ Therefore first one must renormalize the
integrand on the RHS of (\ref{S-4-average-2}).

2) After this step of the renormalization the integrand on the RHS of
(\ref{S-4-average-2}) becomes finite in the internal part of the integration
region but remains singular on the boundary $C$ so that the integral is still
divergent. At this moment we will need the second step of the renormalization.

This two-step renormalization was already announced --- see eq.
(\ref{renormalization-chain}). Note that in the case
of refs.~\cite{Filk-preprint,Dietz-83} both types of divergences were
renormalized using one renormalization procedure based on analytical regularization
(\ref{alpha-sums-DF}) of formal divergent series:\ discrete summation
(\ref{Q-sum-DF}) of the spectral representation keeps \emph{interior} and
\emph{boundary} divergences in one pot. But our method based on
SC\ representation requires a separate treatment of interior and boundary divergences.

Thus we have a two-stage renormalization procedure. A detailed description of
the two steps will be given below. But already now it makes sense to announce
that the second step (renormalization of boundary divergences) will be
implemented via a sort of analytical regularization for divergent integrals in
SC representation. This SC\ version of the analytical regularization (applicable
to arbitrary polygons $C$) is different from the analytical regularization
of refs.~\cite{Filk-preprint,Dietz-83} designed for rectangles and
dealing with discrete series.

\subsection{First step of renormalization:\ from $\left\langle \mathcal{S}
_{4}\right\rangle $ to $\left\langle \mathcal{S}_{4}\right\rangle
^{\text{ren,1}}$}

Function $G_{\mu\nu}$ (\ref{G-mu-nu-def}) is divergent. Indeed, matrix element (\ref{y-x-ME})
has a singularity in the diagonal limit $y\rightarrow x$:
\begin{equation}
\left\langle y\right|  \partial_{\mu}\frac{1}{\Delta\left(  C\right)
}\partial_{\nu}\left|  x\right\rangle \overset{y\rightarrow x}{\sim
}-\frac{\partial}{\partial y^{\mu}}\frac{\partial}{\partial x^{\nu}}
\frac{1}{2\pi}\ln\left|  x-y\right|  =-\frac{r^{2}\delta_{\mu\nu}-2\left(
x-y\right)  _{\mu}\left(  x-y\right)  _{\nu}}{2\pi r^{4}},
\label{propagator-singularity-cartesian}
\end{equation}
\begin{equation}
r=\left|  x-y\right|  .
\end{equation}
This divergence is eliminated by a renormalization of string tension
$\sigma_{0}$. Operationally this renormalization corresponds to the
replacement of $G_{\mu\nu}$  (\ref{G-mu-nu-def}) by its finite part
\begin{equation}
K_{\mu\nu}\left(  x\right)  =\lim_{y\rightarrow x}\left[  \left\langle
y\right|  \partial_{\mu}\frac{1}{-\Delta\left(  C\right)  }\partial_{\nu
}\left|  x\right\rangle -\frac{r^{2}\delta_{\mu\nu}-2\left(  x-y\right)
_{\mu}\left(  x-y\right)  _{\nu}}{2\pi r^{4}}\right]  \label{K-ij-covariant}
\end{equation}
and by the replacement of bare string tension $\sigma_{0}$ with renormalized
string tension $\sigma$ appearing in area law (\ref{Wilson-area-law}).

After the replacement
\begin{equation}
G_{\mu\nu}\rightarrow K_{\mu\nu},
\end{equation}
\begin{equation}
\sigma_{0}\rightarrow\sigma
\end{equation}
eq.~(\ref{S-4-average-2}) takes the form
\begin{equation}
\left\langle \mathcal{S}_{4}\right\rangle ^{\text{ren,1}}=-\frac{D-2}{8\sigma
}\int_{U\left(  C\right)  }d^{2}x\sum\limits_{\mu,\nu=1}^{2}\left[  \left(
4-D\right)  K_{\mu\mu}K_{\nu\nu}+2\left(  D-2\right)  K_{\mu\nu}K_{\mu\nu
}\right]  . \label{S4-average-ren-1-real}
\end{equation}
We use label \emph{ren,1} in $\left\langle \mathcal{S}_{4}\right\rangle
^{\text{ren,1}}$ because the integral on the RHS still contains additional divergences:

\noindent$\bullet$ boundary divergences which appear in $K_{\mu\nu}\left(  x\right)  $ when $x$
approaches boundary $C$ of region $U(C)$,

\noindent$\bullet$ extra cusp divergences appearing in polygonal regions $U\left(  C\right)  $
when $x$ approaches a vertex of the polygon.

Thus, the first step $\left\langle \mathcal{S}_{4}\right\rangle \rightarrow
\left\langle \mathcal{S}_{4}\right\rangle ^{\text{ren,1}}$ of our
renormalization program (\ref{renormalization-chain}) has made the
\emph{integrand} finite in the internal part of $U\left(  C\right)  $ on the
RHS of (\ref{S4-average-ren-1-real}) but the \emph{integral} is still
divergent because of boundary and cusp singularities of the integrand.

\section{Complex representation}

\label{complex-representation-section}

\setcounter{equation}{0} 

\subsection{Definitions and conventions}

In order to proceed, it is convenient to pass from Cartesian 2D coordinates
$x$ to complex coordinates. Our conventions for this complexification are
\begin{align}
\zeta &  =x^{1}+ix^{2},\\
\zeta^{\ast}  &  =x^{1}-ix^{2},
\end{align}
\begin{align}
\frac{\partial}{\partial\zeta}  &  =\frac{1}{2}\left(  \frac{\partial
}{\partial x^{1}}-i\frac{\partial}{\partial x^{2}}\right)  ,\\
\frac{\partial}{\partial\zeta^{\ast}}  &  =\frac{1}{2}\left(  \frac{\partial
}{\partial x^{1}}+i\frac{\partial}{\partial x^{2}}\right)  ,
\end{align}
\begin{equation}
d^{2}\zeta=dx^{1}dx^{2},
\end{equation}
\begin{equation}
ds^{2}=dx^{1}dx^{1}+dx^{2}dx^{2}=d\zeta d\zeta^{\ast}.
\end{equation}
We use the metric tensor with components
\begin{align}
g_{\zeta\zeta^{\ast}}  &  =g_{\zeta^{\ast}\zeta}=\frac{1}{2},\\
g_{\zeta\zeta}  &  =g_{\zeta^{\ast}\zeta^{\ast}}=0,
\end{align}
\begin{align}
g^{\zeta\zeta^{\ast}}  &  =g^{\zeta^{\ast}\zeta}=2,\\
g^{\zeta\zeta}  &  =g^{\zeta^{\ast}\zeta^{\ast}}=0.
\end{align}
For vector $V$
\begin{equation}
V^{\zeta}=2V_{\zeta^{\ast}}=V^{1}+iV^{2},
\end{equation}
\begin{equation}
V^{\zeta^{\ast}}=2V_{\zeta}=V^{1}-iV^{2}.
\end{equation}
For symmetric tensor
\begin{equation}
K_{\mu\nu}=K_{\nu\mu},
\end{equation}
we have
\begin{equation}
K_{\zeta\zeta}=\frac{1}{4}\left(  K_{11}-2iK_{12}-K_{22}\right)  ,
\end{equation}
\begin{equation}
K_{\zeta^{\ast}\zeta^{\ast}}=\frac{1}{4}\left(  K_{11}+2iK_{12}-K_{22}\right)
,
\end{equation}
\begin{equation}
K_{\zeta\zeta^{\ast}}=K_{\zeta^{\ast}\zeta}=\frac{1}{4}\left(  K_{11}
+K_{22}\right)
\end{equation}
so that
\begin{equation}
K_{\mu\nu}K_{\mu\nu}=8\left(  K_{\zeta\zeta^{\ast}}K_{\zeta\zeta^{\ast}
}+K_{\zeta\zeta}K_{\zeta^{\ast}\zeta^{\ast}}\right)  ,
\end{equation}
\begin{equation}
K_{\mu\mu}K_{\nu\nu}=16K_{\zeta\zeta^{\ast}}K_{\zeta\zeta^{\ast}}.
\end{equation}
Now eq.~(\ref{S4-average-ren-1-real}) becomes
\begin{equation}
\left\langle \mathcal{S}_{4}\right\rangle ^{\text{ren,1}}=-\frac{2}{\sigma
}\left(  D-2\right)  \int_{U\left(  C\right)  }d^{2}x\left[  2\left|
K_{\zeta\zeta^{\ast}}\right|  ^{2}+\left(  D-2\right)  \left|  K_{\zeta\zeta
}\right|  ^{2}\right]  . \label{S4-D-ren-complex}
\end{equation}

In real and complex representations for the Green function
\begin{equation}
\left\langle y\right|  \frac{1}{\Delta\left(  C\right)  }\left|
x\right\rangle ,
\end{equation}
we assume the equivalence of various notations
\begin{equation}
\left|  x\right\rangle =\left|  x^{1},x^{2}\right\rangle =\left|  x^{1}
+ix^{2}\right\rangle =\left|  \zeta\right\rangle \,=\left|  \zeta,\zeta^{\ast
}\right\rangle
\end{equation}
representing the same coordinate state.

With the above conventions we can rewrite eq.~(\ref{K-ij-covariant}) in the
complex form:
\begin{equation}
K_{\zeta\zeta}\left(  \zeta\right)  =\lim_{\zeta^{\prime}\rightarrow\zeta
}\left[  \frac{\partial}{\partial\zeta^{\prime}}\frac{\partial}{\partial\zeta
}\left\langle \zeta^{\prime}\right|  \frac{1}{\Delta\left(  C\right)  }\left|
\zeta\right\rangle -\frac{1}{4\pi\left(  \zeta^{\prime}-\zeta\right)  ^{2}
}\right]  , \label{K-zet-zeta-def}
\end{equation}
\begin{equation}
K_{\zeta\zeta^{\ast}}\left(  \zeta\right)  =\lim_{\zeta^{\prime}
\rightarrow\zeta}\left[  \frac{\partial}{\partial\zeta^{\prime}}
\frac{\partial}{\partial\zeta^{\ast}}\left\langle \zeta^{\prime}\right|
\frac{1}{\Delta\left(  C\right)  }\left|  \zeta\right\rangle \right]  .
\label{K-zeta-zeta-bar-def}
\end{equation}

\subsection{Conformal mapping to semiplane}

If one knows a conformal mapping of region $U$ (bounded by Wilson contour $C$)
to the upper complex semiplane then functions $K_{\zeta\zeta},$ $K_{\zeta
\zeta^{\ast}}$ can be expressed via this conformal mapping. The computation is
done in appendix~\ref{Appendix-Diagonal-limit}. The result is

\begin{equation}
K_{\zeta\zeta}\left(  \zeta\right)  =\frac{1}{24\pi}\{z,\zeta\},
\label{K-zeta-zeta-CS}
\end{equation}
\begin{equation}
K_{\zeta\zeta^{\ast}}\left(  \zeta\right)  =\frac{1}{16\pi}\left|
\frac{\partial z\left(  \zeta\right)  }{\partial\zeta}\right|  ^{2}
\frac{1}{\left[  \operatorname{Im}\,z\left(  \zeta\right)  \right]  ^{2}}.
\label{K-zeta-zeta-bar-CS}
\end{equation}
Here $\zeta$ is a complex coordinate in region $U$ and $z\left(  \zeta\right)
$ is the conformal mapping form $U$ to the upper complex $z$-semiplane. We use
notation $\{z,\zeta\}$ for Schwarz derivative (\ref{Schwarz-derivative-def-3}).

\subsection{Reduction of $\left\langle \mathcal{S}_{4}\right\rangle
^{\text{ren,1}}$ to basic integrals $I_{1}$ and $I_{2}$}

Inserting expressions (\ref{K-zeta-zeta-CS}), (\ref{K-zeta-zeta-bar-CS}) into
eq.~(\ref{S4-D-ren-complex}), we obtain

\begin{equation}
\left\langle \mathcal{S}_{4}\right\rangle ^{\text{ren,1}}=-\frac{1}{\sigma
}\frac{D-2}{\left(  8\pi\right)  ^{2}}\left[  I_{1}+\frac{2}{9}\left(
D-2\right)  I_{2}\right] \label{S-ren-1-res}
\end{equation}
where

\begin{align}
I_{1}  &  =\int_{U\left(  C\right)  }d^{2}\zeta\left|  \frac{\partial z\left(
\zeta\right)  }{\partial\zeta}\frac{1}{\left[  \operatorname{Im}\,z\left(
\zeta\right)  \right]  }\right|  ^{4},\label{I1-int-zeta}\\
I_{2}  &  =\int_{U\left(  C\right)  }d^{2}\zeta\left|  \{z,\zeta\}\right|
^{2}. \label{I2-int-zeta}
\end{align}
Next we change the integration variable from $\zeta$ to $z$ in integrals
(\ref{I1-int-zeta}), (\ref{I2-int-zeta}). In expression (\ref{I2-int-zeta})
for $I_{2}$ we use property (\ref{theorem-1-statement}) of Schwarz derivative
\begin{equation}
\{z,\zeta\}=-\left(  \frac{dz}{d\zeta}\right)  ^{2}\{\zeta,z\}.
\end{equation}
Thus
\begin{equation}
I_{1}=\int_{\mathbb{C}_{+}}d^{2}z\left|  \frac{d\zeta}{dz}\right|
^{-2}\left|  \operatorname{Im}\,z\right|  ^{-4}, \label{I-mixed-1}
\end{equation}
\begin{equation}
I_{2}=\int_{\mathbb{C}_{+}}d^{2}z\left|  \frac{d\zeta}{dz}\right|
^{-2}\left|  \left\{  \zeta,z\right\}  \right|  ^{2}. \label{I-holo-1}
\end{equation}
Here the integration runs over upper complex semiplane
\begin{equation}
\mathbb{C}_{+}=\left\{  z:\operatorname{Im}\,z>0\right\}  .
\label{C-2-plus-def}
\end{equation}

\section{Polygons and SC\ mapping}

\label{SC-mapping-section}

\setcounter{equation}{0} 

\subsection{SC\ mapping with finite vertices}

In the case of polygons, conformal mapping $\zeta\left(  z\right)  $ is
known as SC mapping. SC\ mapping $\zeta\left(  z\right)  $ is defined by
differential equation
\begin{equation}
\frac{d\zeta}{dz}=A\prod_{k=1}^{n_{\text{v}}}\left(  z-z_{k}\right)
^{-\beta_{k}}. \label{SC-mapping-general}
\end{equation}
Parameters $\beta_{k}$ determine interior angles of the polygon:
\begin{equation}
\theta_{k}=\pi\left(  1-\beta_{k}\right)  . \label{theta-k-via-beta-k}
\end{equation}
Angles $\theta_{k}$ satisfy geometric conditions
\begin{equation}
0<\theta_{k}<2\pi,\,\theta_{k}\neq\pi,
\end{equation}
\begin{equation}
\sum_{k=1}^{n_{\text{v}}}\left(  \pi-\theta_{k}\right)  =2\pi
\label{angle-sum-rule}
\end{equation}
so that parameters $\beta_{k}$ obey constraints
\begin{equation}
-1<\beta_{k}<1,\,\beta_{k}\neq0, \label{beta-k-range}
\end{equation}
\begin{equation}
\sum_{k=1}^{n_{\text{v}}}\beta_{k}=2. \label{beta-sum-rule}
\end{equation}
Sometimes it is convenient to use parameters
\begin{equation}
\alpha_{k}=1-\beta_{k}=\frac{\theta_{k}}{\pi}.
\label{alpha-k-via-beta-k-theta-k}
\end{equation}

For SC\ mapping (\ref{SC-mapping-general}), Schwarz derivative $\left\{
\zeta,z\right\}  $ (\ref{Schwarz-derivative-def-3}) can be easily computed:
\begin{equation}
\left\{  \zeta,z\right\}  =
\left[
\sum_{k=1}^{n_{\text{v}}}\frac{\beta_{k}}{\left(
z-z_{k}\right)  ^{2}}
\right]
-\frac{1}{2}\left(  \sum_{k=1}^{n_{\text{v}}}
\frac{\beta_{k}}{z-z_{k}}\right)  ^{2}.
\label{Schwarz-derivative-polygon-general}
\end{equation}
Now eqs.~(\ref{I-mixed-1}), (\ref{I-holo-1}) lead to
\begin{equation}
I_{1}=\left|  A\right|  ^{-2}\int_{\mathbb{C}_{+}}d^{2}z\prod
_{k=1}^{n_{\text{v}}}\left|  z-z_{k}\right|  ^{2\beta_{k}}\left|
\operatorname{Im}\,z\right|  ^{-4}, \label{I-mixed-nonreg}
\end{equation}
\begin{equation}
I_{2}=\left|  A\right|  ^{-2}\int_{\mathbb{C}_{+}}d^{2}z\prod
_{k=1}^{n_{\text{v}}}\left|  z-z_{k}\right|  ^{2\beta_{k}}\left|
\left[  \sum
_{k=1}^{n_{\text{v}}}\frac{\beta_{k}}{\left(  z-z_{k}\right)  ^{2}}
\right]
-\frac{1}{2}\left(  \sum_{k=1}^{n_{\text{v}}}\frac{\beta_{k}}{z-z_{k}}\right)
^{2}\right|  ^{2}. \label{I-holo-nonreg-explicit}
\end{equation}
Thus
\begin{equation}
I_{a}=\left|  A\right|  ^{-2}\int_{\mathbb{C}_{+}}d^{2}z\prod
_{k=1}^{n_{\text{v}}}\left|  z-z_{k}\right|  ^{2\beta_{k}}Q_{a}\left(
z,z^{\ast}\right)  \, \label{I-holo-mixed-common-basis}
\end{equation}
where
\begin{align}
Q_{1}\left(  z,z^{\ast}\right)   &  =\left|  \operatorname{Im}\,z\right|
^{-4},\label{Q-mixed-def}\\
Q_{2}\left(  z,z^{\ast}\right)   &  =\left|
\left[  \sum_{k=1}^{n_{\text{v}}
}\frac{\beta_{k}}{\left(  z-z_{k}\right)  ^{2}}
\right]
-\frac{1}{2}\left(  \sum
_{k=1}^{n_{\text{v}}}\frac{\beta_{k}}{z-z_{k}}\right)  ^{2}\right|  ^{2}.
\label{Q-holo-def}
\end{align}

\subsection{SL(2,$\mathbb{R}$) symmetry}

\label{SL-2-R-symmetry-part-1-section}

It is well-known that linear fractional transformations
\begin{equation}
z^{\prime}=\frac{az+b}{cz+d},
\end{equation}
with real coefficients
\begin{equation}
a,b,c,d\in\mathbb{R},
\end{equation}
\begin{equation}
ad-bc=1
\end{equation}
map the upper complex semiplane to itself. The group of these transformations
essentially coincides with SL(2,$\mathbb{R}$) group of matrices
\begin{equation}
\left(
\begin{array}
[c]{cc}
a & b\\
c & d
\end{array}
\right)
\end{equation}
with unit determinant (more exactly, with SL(2,$\mathbb{R}$)/$\mathbb{Z}_{2}$).

For a given polygon its SC\ mapping is defined up to the freedom of these
linear fractional transformations. In particular, this allows for the freedom
of changing SC\ vertices $z_{k}$
\begin{equation}
z_{k}^{\prime}=\frac{az_{k}+b}{cz_{k}+d}. \label{SL-2-R-z-k}
\end{equation}
Naive expressions (\ref{Q-mixed-def}), (\ref{Q-holo-def}) are formally
invariant under this symmetry. But when it comes to the renormalization of
ultraviolet divergences in naive integrals, this SL(2,$\mathbb{R}$) symmetry
imposes nontrivial constraints on the renormalization procedure.

\subsection{SC\ mapping with one vertex at infinity}

The freedom of linear fractional transformations (\ref{SL-2-R-z-k}) allows for
taking one SC\ vertex to infinity. Let us put vertex $z_{n_{\text{v}}}$ to
infinity:
\begin{equation}
\left|  z_{n_{\text{v}}}\right|  \rightarrow\infty.
\label{z-n-v-limit-to-infinity}
\end{equation}
When taking this limit, one should tune parameter $A=A\left(  z_{n_{\text{v}}
}\right)  $ in eq.~(\ref{SC-mapping-general}) so that
\begin{equation}
\lim_{\left|  z_{n_{\text{v}}}\right|  \rightarrow\infty}A\left(
z_{n_{\text{v}}}\right)  \left(  z-z_{n_{\text{v}}}\right)  ^{-\beta
_{n_{\text{v}}}}=\lim_{\left|  z_{n_{\text{v}}}\right|  \rightarrow\infty
}A\left(  z_{n_{\text{v}}}\right)  \left(  -z_{n_{\text{v}}}\right)
^{-\beta_{n_{\text{v}}}}=\tilde{A}=\text{finite}\neq0.
\label{A-tilde-A-condition}
\end{equation}
Then SC\ equation (\ref{SC-mapping-general}) reduces to

\begin{equation}
\frac{d\zeta}{dz}=\tilde{A}\prod_{k=1}^{n_{\text{v}}-1}\left(  z-z_{k}\right)
^{-\beta_{k}}. \label{SC-tilded-1}
\end{equation}
This equation does not contain parameter $\beta_{n_{\text{v}}}$ but it still
can be defined by eq.~(\ref{beta-sum-rule})

\begin{equation}
\beta_{n_{\text{v}}}=2-\sum_{k=1}^{n_{\text{v}}-1}\beta_{k}
\label{beta-n-v-via-other}
\end{equation}
and obeys condition (\ref{beta-k-range}) together with other parameters
$\beta_{k}$:
\begin{equation}
-1<\beta_{k}<1\quad(1\leq k\leq n_{\text{v}}).
\label{beta-k-constraints-SC-vertex-infinite}
\end{equation}
The expression for Schwarz derivative
(\ref{Schwarz-derivative-polygon-general}) of mapping (\ref{SC-tilded-1}) can
be obtained by taking limit (\ref{z-n-v-limit-to-infinity})
in eq. ~(\ref{Schwarz-derivative-polygon-general}):
\begin{equation}
\left\{  \zeta,z\right\}  =\left[\sum_{k=1}^{n_{\text{v}}-1}\frac{\beta_{k}}{\left(
z-z_{k}\right)  ^{2}}\right]
-\frac{1}{2}\left(  \sum_{k=1}^{n_{\text{v}}
-1}\frac{\beta_{k}}{z-z_{k}}\right)  ^{2}.
\end{equation}

Now we apply general formulas (\ref{I-holo-mixed-common-basis}) --
(\ref{Q-mixed-def}) to SC\ mapping (\ref{SC-tilded-1}):
\begin{equation}
I_{a}=\left|  \tilde{A}\right|  ^{-2}\int_{\mathbb{C}_{+}}d^{2}
z\prod_{k=1}^{n_{\text{v}}-1}\left|  z-z_{k}\right|  ^{2\beta_{k}}\tilde
{Q}_{a}\left(  z,z^{\ast}\right)  . \label{I-holo-mixed-common-tilded}
\end{equation}
Here
\begin{align}
\tilde{Q}_{1}\left(  z,z^{\ast}\right)   &  =\left|  \operatorname{Im}
\,z\right|  ^{-4},\label{Q-mixed-def-2}\\
\tilde{Q}_{2}\left(  z,z^{\ast}\right)   &  =\left|
\left[
  \sum_{k=1}^{n_{\text{v}
}-1}\frac{\beta_{k}}{\left(  z-z_{k}\right)  ^{2}}
\right]
-\frac{1}{2}\left(
\sum_{k=1}^{n_{\text{v}}-1}\frac{\beta_{k}}{z-z_{k}}\right)  ^{2}\right|
^{2}. \label{Q-holo-def-2}
\end{align}

\subsection{SC\ vertex at infinity: pro and contra}

\label{pro-contra-section}

Thus we have two versions of SC\ representation for $I_{a}$. Representation
(\ref{I-holo-mixed-common-basis}) is based on SC\ mapping with finite vertices
$z_{k}$ whereas representation (\ref{I-holo-mixed-common-tilded}) assumes that
SC\ vertex $z_{n_{\text{v}}}$ is taken to infinity. Both integrals
(\ref{I-holo-mixed-common-basis}), (\ref{I-holo-mixed-common-tilded}) are
divergent. These integrals must be renormalized (by the second step
$\left\langle \mathcal{S}_{4}\right\rangle ^{\text{ren,1}}\rightarrow
\left\langle \mathcal{S}_{4}\right\rangle ^{\text{ren}}$ of our
renormalization program (\ref{renormalization-chain})) and our plan is to use
analytical regularization. In principle, our procedure of analytical
regularization can be formulated for both versions of SC\ mapping with the
same final renormalized result for $I_{a}^{\text{ren}}$. But when it comes to
the practical work, the SC\ representation with one vertex at infinity seems
to be preferable. One advantage of this approach is obvious: keeping one
SC\ vertex at infinity we significantly simplify all integrals. The second
argument against SC mapping with finite vertices is that parameters $\beta
_{k}$ must obey constraint (\ref{beta-sum-rule}). This constraint is critical
at some steps of analytical regularization so that one meets the problem
of analytical continuation in $\beta_{k}$ under constraint
(\ref{beta-sum-rule}) which leads to some technical (solvable but annoying) problems.

As was discussed in section~\ref{SL-2-R-symmetry-part-1-section}, in the case of
SC\ mapping with all SC\ vertices being finite we have a powerful constraint
of SL(2,$\mathbb{R}$) symmetry on allowed renormalization procedures. Our
choice to construct renormalization in terms of SC\ mapping with one vertex at
infinity makes SL$\left(  2,\mathbb{R}\right)  $ symmetry rather implicit. We
are left with the freedom of choice: \emph{which} vertex of the polygon is
associated with the SC\ vertex taken to infinity. The result of the
renormalization must be independent of this choice.

\section{Analytical regularization}

\label{Analytical-regularization-section}

\setcounter{equation}{0} 

\subsection{Functions $M_{P}^{(n)}$}

Divergent integrals (\ref{I-holo-mixed-common-tilded}) representing $I_{1}$,
$I_{2}$ belong to the class of integrals
\begin{equation}
M_{P}^{(n)\text{nonren}}\left(  \left\{  \gamma_{k}\right\}  _{k=0}
^{n},\left\{  z_{k}\right\}  _{k=1}^{n}\right)  =\int_{\mathbb{C}_{+}
}d^{2}z\,\left|  \operatorname{Im}\,z\right|  ^{\gamma_{0}}\prod_{k=1}
^{n}\left|  z-z_{k}\right|  ^{\gamma_{k}}P\left(  z,z^{\ast}\right)
\label{M-P-int-naive-def}
\end{equation}
where
\begin{equation}
n\geq2,
\end{equation}
$P\left(  z,z^{\ast}\right)  $ is a polynomial and $z_{k}$ ($1\leq k\leq n$) are
different real numbers:
\begin{equation}
z_{j}\neq z_{k}\,\text{if\thinspace}j\neq k\,.
\end{equation}
We use
label \emph{nonren} in $M_{P}^{(n)\text{nonren}}$ because the integral on the
RHS of eq.~(\ref{M-P-int-naive-def}) may be divergent.

Expression (\ref{I-holo-mixed-common-tilded}) can be rewritten for $I_{1}$ in
the form
\begin{align}
I_{1}  &  =\left|  \tilde{A}\right|  ^{-2}\int_{\mathbb{C}_{+}}d^{2}
z\prod_{k=1}^{n_{\text{v}}-1}\left|  z-z_{k}\right|  ^{2\beta_{k}}\left|
\operatorname{Im}\,z\right|  ^{-4}\nonumber\\
&  =M_{1}^{(n_{\text{v}}-1)\text{nonren}}\left(  -4,\left\{  2\beta
_{k}\right\}  _{k=1}^{n_{\text{v}}-1},\left\{  z_{k}\right\}  _{k=1}
^{n_{\text{v}}-1}\right)  \label{I-mixed-M-1}
\end{align}
where $M_{1}^{(n)\text{nonren}}$ is $M_{P}^{(n)\text{nonren}}$ with trivial
constant polynomial $P\left(  z,z^{\ast}\right)  \equiv1$.

For $I_{2}$ we find from eqs.~(\ref{I-holo-mixed-common-tilded}) and (\ref{Q-holo-def-2})
\begin{align}
I_{2}  &  =\left|  \tilde{A}\right|  ^{-2}\int_{\mathbb{C}_{+}}d^{2}
z\prod_{k=1}^{n_{\text{v}}-1}\left|  z-z_{k}\right|  ^{2\beta_{k}}\left|
\left[
\sum_{k=1}^{n_{\text{v}}-1}\frac{\beta_{k}}{\left(  z-z_{k}\right)  ^{2}}
\right]
-\frac{1}{2}\left(  \sum_{k=1}^{n_{\text{v}}-1}\frac{\beta_{k}}{z-z_{k}
}\right)  ^{2}\right|  ^{2}\nonumber\\
&  =\left|  \tilde{A}\right|  ^{-2}\int_{\mathbb{C}_{+}}d^{2}z\prod
_{k=1}^{n_{\text{v}}-1}\left|  z-z_{k}\right|  ^{2\beta_{k}-4}P_{2}\left(
z,z^{\ast}\right) \nonumber\\
&  =M_{P_{2}}^{(n_{\text{v}}-1)\text{nonren}}\left(  0,\left\{  2\beta
_{k}-4\right\}  _{k=1}^{n_{\text{v}}-1},\left\{  z_{k}\right\}  _{k=1}
^{n_{\text{v}}-1}\right)  \label{I-holo-M-P}
\end{align}
where polynomial
\begin{equation}
P_{2}\left(  z,z^{\ast}\right)  =T_{2}\left(  z\right)  \left[  T_{2}\left(
z\right)  \right]  ^{\ast} \label{P-holo-factorized}
\end{equation}
is defined via holomorphic polynomial
\begin{equation}
T_{2}\left(  z\right)  =\left\{  \left[  \sum_{k=1}^{n_{\text{v}}
-1}\frac{\beta_{k}}{\left(  z-z_{k}\right)  ^{2}}\right]  -\frac{1}{2}\left(
\sum_{k=1}^{n_{\text{v}}-1}\frac{\beta_{k}}{z-z_{k}}\right) ^{2}\right\}
\prod_{k=1}^{n_{\text{v}}-1}\left(  z-z_{k}\right)  ^{2}. \label{R-holo-def}
\end{equation}

One can construct the analytical regularization and renormalization of integrals
(\ref{M-P-int-naive-def}) by analogy with the theory of Euler B-function
\begin{equation}
B\left(  \gamma_{1},\gamma_{2}\right)  =\int_{0}^{1}dx\,x^{\gamma_{1}
-1}\left(  1-x\right)  ^{\gamma_{2}-1}.
\end{equation}
First we define functions $M_{P}^{(n)}\left(  \left\{  \gamma
_{k}\right\}  _{k=0}^{n},\left\{  z_{k}\right\}  _{k=1}^{n}\right)  $ by
integrals (\ref{M-P-int-naive-def}) in the region of parameters $\left\{  \gamma
_{k}\right\}  _{k=0}^{n}$ where these integrals are convergent and then
perform analytical continuation in variables $\left\{  \gamma_{k}\right\}
_{k=0}^{n}$ to physical values appearing in eqs.~(\ref{I-mixed-M-1}),
(\ref{I-holo-M-P}). This analytical continuation will be denoted as
\begin{eqnarray}
&&M_{P}^{(n)}\left( \left\{ \gamma _{k}\right\} _{k=0}^{n},\left\{
z_{k}\right\} _{k=1}^{n}\right)   \nonumber \\
&=&\text{analyt. cont. in }\gamma _{k}\left[ \int_{\mathbb{C}
_{+}}d^{2}z\,\left| \operatorname{Im}\,z\right| ^{\gamma _{0}}\prod_{k=1}^{n}\left|
z-z_{k}\right| ^{\gamma _{k}}P\left( z,z^{\ast }\right) \right] \,.
\label{M-P-int-def}
\end{eqnarray}
The details of the analytical continuation are discussed
in section~\ref{Details-analytical-continuation-section}.

Once analytical function $M_{P}^{(n)}\left(  \left\{  \gamma_{k}\right\}
_{k=0}^{n},\left\{  z_{k}\right\}  _{k=1}^{n}\right)  $ is constructed, one
can use it to define analytical renormalization of $I_{1}$ and $I_{2}$ by
replacing $M_{P}^{(n)\text{nonren}}\rightarrow M_{P}^{(n)}$ in naive divergent integrals
(\ref{I-mixed-M-1}), (\ref{I-holo-M-P})
\begin{equation}
I_{1}^{\text{ren}}=\left|  \tilde{A}\right|  ^{-2}M_{1}^{(n_{\text{v}}
-1)}\left(  -4,\left\{  2\beta_{k}\right\}  _{k=1}^{n_{\text{v}}-1},\left\{
z_{k}\right\}  _{k=1}^{n_{\text{v}}-1}\right)  , \label{I-ren-mixed-via-M}
\end{equation}
\begin{equation}
I_{2}^{\text{ren}}=\left|  \tilde{A}\right|  ^{-2}M_{P_{2}}^{(n_{\text{v}}
-1)}\left(  0,\left\{  2\beta_{k}-4\right\}  _{k=1}^{n_{\text{v}}-1},\left\{
z_{k}\right\}  _{k=1}^{n_{\text{v}}-1}\right)  . \label{I-ren-holo-via-M}
\end{equation}

These values of $I_{1}^{\text{ren}}$, $I_{2}^{\text{ren}}$ must be used in the
completely renormalized version of eq.~(\ref{S-ren-1-res})
\begin{equation}
\left\langle \mathcal{S}_{4}\right\rangle ^{\text{ren}}=-\frac{1}{\sigma
}\frac{D-2}{\left(  8\pi\right)  ^{2}}\left[  I_{1}^{\text{ren}}+\frac{2}
{9}\left(  D-2\right)  I_{2}^{\text{ren}}\right]  . \label{S-ren-via-I-ren}
\end{equation}
The transition to eqs.~(\ref{I-ren-mixed-via-M}), (\ref{I-ren-holo-via-M}),
(\ref{S-ren-via-I-ren}) corresponds to the second step $\left\langle
\mathcal{S}_{4}\right\rangle ^{\text{ren,1}}\rightarrow\left\langle
\mathcal{S}_{4}\right\rangle ^{\text{ren}}$ of our renormalization program
(\ref{renormalization-chain}).

\subsection{From divergent integrals to meromorphic functions}

\label{Details-analytical-continuation-section}

The suggested scheme of analytical renormalization raises many questions if
one wants a mathematically impeccable implementation of this program. On the
other hand, in practical calculations one can typically (but not always)
perform formal mathematical operations with divergent integrals without
wasting time for a careful justification of these naive formal manipulations.
A careful mathematical theory of functions $M_{P}^{(n)}$ for arbitrary $n$ and
arbitrary polynomials $P$ can be constructed. But in the framework of the
current work devoted to triangular Wilson loops we need functions $M_{P}^{(n)}$
only for the special case $n=2$ and for rather simple polynomials $P$. In
this simple case the integrals can be computed explicitly and
the problem of analytical continuation can be solved by using
these explicit expressions rather than invoking the general theory of
functions $M_{P}^{(n)}$ with arbitrary $n$ and arbitrary $P$.

Therefore the complete description of the theory of functions $M_{P}^{(n)}$
will be presented in a separate work devoted to Wilson loops with arbitrary
polygonal contours. Here only a list of main results will be given without
proofs and derivations.

When one starts with the construction of the rigorous theory of functions
$M_{P}^{(n)}$ one meets several questions:

1) Is the region of parameters $\left\{  \gamma_{k}\right\}  _{k=0}^{n}$ where
integral (\ref{M-P-int-def}) is convergent non-empty? In other words, do we
have a starting region for the construction of the analytical continuation?

2) Does the analytical continuation depend on the path?

3) Is the result of the analytical continuation regular at final physical
points appearing on the RHS\ of eqs.~(\ref{I-ren-mixed-via-M}),
(\ref{I-ren-holo-via-M})?

The brief answers are:

1) The region of parameters $\left\{  \gamma_{k}\right\}  _{k=0}^{n}$ where
integral (\ref{M-P-int-def}) is convergent is non-empty for any fixed
$\left\{  z_{k}\right\}  _{k=1}^{n}$ and for any fixed $P$.

2) Starting from function $M_{P}^{(n)}\left(  \left\{  \gamma
_{k}\right\}  _{k=0}^{n},\left\{  z_{k}\right\}  _{k=1}^{n}\right)  $ defined
by integral (\ref{M-P-int-def}) in the convergence region and performing
analytical continuation in $\mathbb{C}^{n+1}$ space of parameters $\left\{
\gamma_{k}\right\}  _{k=0}^{n}$ (at fixed $\left\{  z_{k}\right\}  _{k=1}^{n}$
and fixed $P$), one arrives at meromorphic function $M_{P}^{(n)}$ 
of $\left\{\gamma_{k}\right\}  _{k=0}^{n}$
(in the sense of theory of several complex variables). In practical terms this means
that analytical continuation of $M_{P}^{(n)}$ is independent of
the path in the multidimensional space of complex $\left\{  \gamma
_{k}\right\}  _{k=0}^{n}$ but one can meet pole singularities.

3) One can show that $\left\{  \gamma_{k}\right\}  _{k=0}^{n}$ arguments of
$M_{P}^{(n)}$ appearing on the RHS\ of eqs.~(\ref{I-ren-mixed-via-M}),
(\ref{I-ren-holo-via-M}) are regular points of meromorphic function
$M_{P}^{(n)}$ (if parameters $\beta_k$ satisfy geometric constraint (\ref{beta-k-constraints-SC-vertex-infinite})). In other words, expressions (\ref{I-ren-mixed-via-M}),
(\ref{I-ren-holo-via-M}) give finite and unambiguous expressions for
$I_{1}^{\text{ren}}$ and $I_{2}^{\text{ren}}$.

A careful proof of these statements for arbitrary $n\geq2$ (i.e. for polygons
with arbitrary number of vertices $n_{\text{v}}\geq3$) requires some effort. We
postpone the proof till a separate work devoted to arbitrary polygons. In the
case of triangles, i.e. for $n=n_{\text{v}}-1=2$ many of the above general
properties can be seen directly from explicit expressions computed
in appendices~\ref{Appendix-K-1-2-calculation}, \ref{Appendix-K-P-2-calculation}.

\subsection{Functions $\Pi_{P}^{(n)}$}

\label{functions-Pi-section}

The above discussion of the analytical regularization proceeded in terms of
functions $M_{P}^{(n)}$. These functions are convenient for the preliminary
description of our analytical regularization. But when it comes to the real
work (including proofs of the announced statements and the practical calculation of
$f_{2}\left(  C\right)  $ for various polygons) then it makes sense to use
other basic functions which differ from $M_{P}^{(n)}\left(  \left\{
\gamma_{k}\right\}  _{k=0}^{n},\left\{  z_{k}\right\}  _{k=1}^{n}\right)  $ by
a simple linear change of arguments. The new functions are defined by the
relation
\begin{equation}
\Pi_{P}^{(n)}\left(  \alpha,\left\{  \gamma_{k}\right\}  _{k=1}^{n},\left\{
z_{k}\right\}  _{k=1}^{n}\right)  =M_{P}^{(n)}\left(  \alpha-1,\left\{
\gamma_{k}-\alpha-1\right\}  _{k=1}^{n},\left\{  z_{k}\right\}  _{k=1}
^{n}\right)  \,
\end{equation}
with the inverse expression
\begin{equation}
M_{P}^{(n)}\left(  \gamma_{0},\left\{  \gamma_{k}\right\}  _{k=1}^{n},\left\{
z_{k}\right\}  _{k=1}^{n}\right)  \,=\Pi_{P}^{(n)}\left(  \gamma
_{0}+1,\left\{  \gamma_{k}+\gamma_{0}+2\right\}  _{k=1}^{n},\left\{
z_{k}\right\}  _{k=1}^{n}\right)  . \label{M-P-via-K-P}
\end{equation}

Using eq.~(\ref{M-P-int-def}), we find that
functions $\Pi_{P}^{(n)}$ are given by analytical continuation in $\mathbb{C}^{n+1}$ space of parameters
$\alpha$,  $\left\{\gamma_{k}\right\}_{k=1}^{n}$
\begin{align}
&  \Pi_{P}^{(n)}\left(  \alpha,\left\{  \gamma_{k}\right\}_{k=1}^{n},\left\{  z_{k}\right\} _{k=1}^{n}\right)
\nonumber\\
&  =\text{analyt. cont. in }\alpha, \gamma _{k}
\left[ \int_{\mathbb{C}_{+}}d^{2}z\,\left|  \operatorname{Im}\,z\right|
^{\alpha-1}\prod_{k=1}^{n}\left|  z-z_{k}\right|  ^{\gamma_{k}-\alpha
-1}P\left(  z,z^{\ast}\right) \right] . \label{K-P-def-integral-naive-1}
\end{align}

In terms of functions $\Pi_{P}^{(n)}$ relations (\ref{I-ren-mixed-via-M}),
(\ref{I-ren-holo-via-M}) take the form
\begin{equation}
I_{1}^{\text{ren}}=\left|  \tilde{A}\right|  ^{-2}\Pi_{1}^{(n_{\text{v}}
-1)}\left(  -3,\left\{  2\beta_{k}-2\right\}  _{k=1}^{n_{\text{v}}-1},\left\{
z_{k}\right\}  _{k=1}^{n_{\text{v}}-1}\right)  , \label{I-ren-mixed-via-K}
\end{equation}
\begin{equation}
I_{2}^{\text{ren}}=\left|  \tilde{A}\right|  ^{-2}\Pi_{P_{2}}^{(n_{\text{v}
}-1)}\left(  1,\left\{  2\beta_{k}-2\right\}  _{k=1}^{n_{\text{v}}-1},\left\{
z_{k}\right\}  _{k=1}^{n_{\text{v}}-1}\right)  . \label{I-ren-holo-via-K}
\end{equation}

Functions $\Pi_{P}^{(n)}$ inherit properties of functions $M_{P}^{(n)}$ discussed
in section~\ref{Details-analytical-continuation-section}. In particular, functions
$\Pi_{P}^{(n)}\left(  \alpha,\left\{  \gamma_{k}\right\}_{k=1}^{n},\left\{  z_{k}\right\} _{k=1}^{n}\right)$
are meromorphic functions in the $\mathbb{C}^{n+1}$ space of complex parameters $\alpha,\left\{  \gamma_{k}\right\}_{k=1}^{n}$
and are regular at values appearing on the RHS of eqs. (\ref{I-ren-mixed-via-K}), (\ref{I-ren-holo-via-K})
if geometric condition (\ref{beta-k-constraints-SC-vertex-infinite}) holds.

\section{Calculation of $f_{2}\left(  C\right)  $ for triangles}

\label{case-of-triangles-section}

\setcounter{equation}{0} 

\subsection{Case of triangles}

In the case of triangles eqs.~(\ref{I-ren-mixed-via-K}),
(\ref{I-ren-holo-via-K}) take the form
\begin{equation}
I_{1}^{\text{ren}}=\left|  \tilde{A}\right|  ^{-2}\Pi_{1}^{(2)}\left(
-3,\left\{  2\beta_{1}-2,2\beta_{2}-2\right\}  ,\left\{  z_{1},z_{2}\right\}
\right)  , \label{I-ren-mixed-triangle}
\end{equation}
\begin{equation}
I_{2}^{\text{ren}}=\left|  \tilde{A}\right|  ^{-2}\Pi_{P_{2}}^{(2)}\left(
1,\left\{  2\beta_{1}-2,2\beta_{2}-2\right\}  ,\left\{  z_{1},z_{2}\right\}
\right)  . \label{I-ren-holo-triangle}
\end{equation}
Below we will see that function $\Pi_{1}^{(2)}\left(  \alpha,\left\{
\gamma_{1},\gamma_{2}\right\}  ,\left\{  z_{1},z_{2}\right\}  \right)  $ with
arbitrary arguments can be easily computed analytically, which will
immediately lead to the result for $I_{1}^{\text{ren}}$. As for $I_{2}
^{\text{ren}}$, one can compute $\Pi_{P}^{(2)}\left(  \alpha,\left\{
\gamma_{1},\gamma_{2}\right\}  ,\left\{  z_{1},z_{2}\right\}  \right)  $ on
the subspace $\alpha=1$ with arbitrary $\gamma_{1},\gamma_{2}$, which is
sufficient for the computation of $I_{2}^{\text{ren}}$
(\ref{I-ren-holo-triangle}).

Without losing generality we perform the computation of $I_{1}^{\text{ren}}$
and $I_{2}^{\text{ren}}$ choosing SC\ vertices
\begin{equation}
z_{1}=0,\quad z_{2}=1. \label{z1-z2-stndard-0}
\end{equation}

\subsection{Calculation of $I_{1}^{\text{ren}}$}

In appendix~\ref{Appendix-K-1-2-calculation} we derive expression
(\ref{K-1-2-res}) for function $\Pi_{1}^{(2)}$. Using this result in our
expression (\ref{I-ren-mixed-triangle}) for $I_{1}^{\text{ren}}$, we find
\begin{equation}
I_{1}^{\text{ren}}=\left|  \tilde{A}\right|  ^{-2}\frac{2\pi}{3}
\frac{\Gamma\left(  \beta_{1}-1\right)  }{\Gamma\left(  -\beta_{1}\right)
}\frac{\Gamma\left(  \beta_{2}-1\right)  }{\Gamma\left(  -\beta_{2}\right)
}\frac{\Gamma\left(  1-\beta_{1}-\beta_{2}\right)  }{\Gamma\left(  \beta
_{1}+\beta_{2}-2\right)  }.
\end{equation}
Next we express parameter $\left|  \tilde{A}\right|  $ via area $S\left(
C\right)  $ of the triangle using eq.~(\ref{A-tilde-via-S-infty}) and simplify
the result using expression (\ref{beta-3-via-beta-1-beta-2}) for $\beta_{3}$ :
\begin{align}
I_{1}^{\text{ren}}  &  =\frac{\pi^{2}}{3S\left(  C\right)  }\left[
\frac{\Gamma\left(  1-\beta_{1}\right)  \Gamma\left(  1-\beta_{2}\right)
\Gamma\left(  1-\beta_{3}\right)  }{\Gamma\left(  \beta_{1}\right)
\Gamma\left(  \beta_{2}\right)  \Gamma\left(  \beta_{3}\right)  }\right]
\nonumber\\
&  \times\left[  \frac{\Gamma\left(  \beta_{1}-1\right)  }{\Gamma\left(
-\beta_{1}\right)  }\frac{\Gamma\left(  \beta_{2}-1\right)  }{\Gamma\left(
-\beta_{2}\right)  }\frac{\Gamma\left(  \beta_{3}-1\right)  }{\Gamma\left(
-\beta_{3}\right)  }\right] \nonumber\\
&  =\frac{\pi^{2}}{3S\left(  C\right)  }\frac{\beta_{1}}{1-\beta_{1}
}\frac{\beta_{2}}{1-\beta_{2}}\frac{\beta_{3}}{1-\beta_{3}}.
\end{align}
Thus
\begin{equation}
I_{1}^{\text{ren}}=\frac{\pi^{2}}{3S\left(  C\right)  }\prod_{k=1}
^{3}\frac{\beta_{k}}{1-\beta_{k}}. \label{I-1-ren-triangle-res}
\end{equation}

\subsection{Calculation of $I_{2}^{\text{ren}}$}

\label{I2-ren-calc}

Now we turn to the computation of $I_{2}^{\text{ren}}$,
 using its expression
(\ref{I-ren-holo-triangle}) via $\Pi_{P_{2}}^{(2)}$. Remember that polynomial
$P_{2}$ can be expressed via polynomial $T_{2}$ according to eq.
(\ref{P-holo-factorized}). Let us compute polynomial $T_{2}\left(  z\right)  $
(\ref{P-holo-factorized}) for triangles ($n_{\text{v}}=3$) with $z_{k}$ given
by eq.~(\ref{z1-z2-stndard-0}). We have
\begin{align}
T_{2}\left(  z\right)   &  =\left\{
\left[
 \sum_{k=1}^{2}\frac{\beta_{k}}{\left(
z-z_{k}\right)  ^{2}}
\right]
-\frac{1}{2}\left(  \sum_{k=1}^{2}\frac{\beta_{k}
}{z-z_{k}}\right)  ^{2}\right\}  \prod_{k=1}^{2}\left(  z-z_{k}\right)
^{2}\nonumber\\
&  =\left\{  \frac{\beta_{1}}{z^{2}}+\frac{\beta_{2}}{\left(  z-1\right)
^{2}}-\frac{1}{2}\left(  \frac{\beta_{1}}{z}+\frac{\beta_{2}}{z-1}\right)
^{2}\right\}  z^{2}\left(  z-1\right)  ^{2}\nonumber\\
&  =\left[  \beta_{1}\left(  1-\frac{1}{2}\beta_{1}\right)  +\beta_{2}\left(
1-\frac{1}{2}\beta_{2}\right)  -\beta_{1}\beta_{2}\right]  z^{2}\nonumber\\
&  +\left[  -2\beta_{1}+\beta_{1}\left(  \beta_{1}+\beta_{2}\right)  \right]
z+\beta_{1}\left(  1-\frac{1}{2}\beta_{1}\right)  .
\end{align}
This can be rewritten as
\begin{equation}
T_{2}\left(  z\right)  =\sum_{n=0}^{2}d_{n}z^{n},
\end{equation}
where
\begin{align}
d_{0}  &  =\beta_{1}\left(  1-\frac{1}{2}\beta_{1}\right)  ,\label{d-0-res}\\
d_{1}  &  =-2\beta_{1}+\beta_{1}\left(  \beta_{1}+\beta_{2}\right)
,\label{d-1-res}\\
d_{2}  &  =\left[  \beta_{1}\left(  1-\frac{1}{2}\beta_{1}\right)  +\beta
_{2}\left(  1-\frac{1}{2}\beta_{2}\right)  -\beta_{1}\beta_{2}\right]  .
\label{d-2-res}
\end{align}
Now eq.~(\ref{P-holo-factorized}) takes the form
\begin{equation}
P_{2}\left(  z,z^{\ast}\right)  =T_{2}\left(  z\right)  \left[  T_{2}\left(
z\right)  \right]  ^{\ast}=\left(  \sum_{m=0}^{2}d_{n}z^{n}\right)  \left(
\sum_{n=0}^{2}d_{n}z^{\ast n}\right)  .
\end{equation}
Applying expression (\ref{K-P-2-generl-res}) for $\Pi_{P}^{(2)\text{ }}\left(
1,\left\{  \gamma_{1},\gamma_{2}\right\}  ,\left\{  0,1\right\}  \right)  $
derived in appendix~\ref{Appendix-K-P-2-calculation} to our case, we find

\begin{align}
\Pi_{P_{2}}^{(2)\text{ }}\left(  1,\left\{  \gamma_{1},\gamma_{2}\right\}
,\left\{  0,1\right\}  \right)   &  =\frac{\pi}{2}\frac{\sin\left(
\pi\frac{\gamma_{1}}{2}\right)  }{\sin\left(  \pi\frac{\gamma_{1}+\gamma_{2}
}{2}\right)  }\frac{\Gamma\left(  \frac{\gamma_{2}}{2}\right)  }{\Gamma\left(
1-\frac{\gamma_{2}}{2}\right)  }\nonumber\\
&  \times\left[  \sum_{m}d_{m}\frac{\Gamma\left(  \frac{\gamma_{1}}
{2}+m\right)  }{\Gamma\left(  \frac{\gamma_{1}+\gamma_{2}}{2}+m\right)
}\right]  ^{2}. \label{K2-P2-calc-3}
\end{align}
In eq.~(\ref{I-ren-holo-via-K}) we are interested in the case

\begin{equation}
\gamma_{k}=2\beta_{k}-2
\end{equation}
where eq.~(\ref{K2-P2-calc-3}) gives
\begin{align}
\Pi_{P_{2}}^{(2)\text{ }}\left(  1,\left\{  2\beta_{1}-2,2\beta_{2}-2\right\}
,\left\{  0,1\right\}  \right)   &  =-\frac{\pi}{2}\frac{\sin\left(  \pi
\beta_{1}\right)  }{\sin\left(  \pi\left(  \beta_{1}+\beta_{2}\right)
\right)  }\frac{\Gamma\left(  \beta_{2}-1\right)  }{\Gamma\left(  2-\beta
_{2}\right)  }\nonumber\\
&  \times\left[  \sum_{m}d_{m}\frac{\Gamma\left(  \beta_{1}-1+m\right)
}{\Gamma\left(  \beta_{1}+\beta_{2}-2+m\right)  }\right]  ^{2}.
\label{K2-P2-calc-4}
\end{align}
Using eqs.~(\ref{d-0-res}), (\ref{d-1-res}), (\ref{d-2-res}), we compute
\begin{equation}
\sum_{m=0}^{2}d_{m}\frac{\Gamma\left(  \beta_{1}+m\right)  }{\Gamma\left(
\beta_{1}+\beta_{2}+m\right)  }=\frac{1}{2}\frac{\Gamma\left(  \beta
_{1}-1\right)  }{\Gamma\left(  \beta_{1}+\beta_{2}\right)  }\beta_{1}\beta
_{2}\left(  \beta_{1}+\beta_{2}-2\right)
\end{equation}
and insert this into eq.~(\ref{K2-P2-calc-4})
\begin{align}
&  \Pi_{P_{2}}^{(2)\text{ }}\left(  1,\left\{  2\beta_{1}-2,2\beta
_{2}-2\right\}  ,\left\{  0,1\right\}  \right) \nonumber\\
&  =-\frac{\pi}{8}\frac{\sin\left(  \pi\beta_{1}\right)  }{\sin\left(
\pi\left(  \beta_{1}+\beta_{2}\right)  \right)  }\frac{\Gamma\left(  \beta
_{2}-1\right)  }{\Gamma\left(  2-\beta_{2}\right)  }\left[  \frac{\Gamma
\left(  \beta_{1}-1\right)  }{\Gamma\left(  \beta_{1}+\beta_{2}\right)  }
\beta_{1}\beta_{2}\left(  \beta_{1}+\beta_{2}-2\right)  \right]  ^{2}.
\label{K2-P2-calc-5}
\end{align}
We introduce according to eq.~(\ref{beta-n-v-via-other})
\begin{equation}
\beta_{3}=2-\beta_{1}-\beta_{2}
\end{equation}
and simplify eq.~(\ref{K2-P2-calc-5})
\begin{equation}
\Pi_{P_{2}}^{(2)\text{ }}\left(  1,\left\{  2\beta_{1}-2,2\beta_{2}-2\right\}
,\left\{  0,1\right\}  \right)  =\frac{\pi}{8}\prod_{k=1}^{3}\left[
\frac{\beta_{k}^{2}\Gamma\left(  \beta_{k}-1\right)  }{\Gamma\left(
2-\beta_{k}\right)  }\right]  .
\end{equation}
We insert this into eq.~(\ref{I-ren-holo-via-K})
\begin{equation}
I_{2}^{\text{ren}}=\left|  \tilde{A}\right|  ^{-2}\frac{\pi}{8}\prod_{k=1}
^{3}\left[  \frac{\beta_{k}^{2}\Gamma\left(  \beta_{k}-1\right)  }
{\Gamma\left(  2-\beta_{k}\right)  }\right]  .
\end{equation}
Now we insert $\left|  \tilde{A}\right|  $ from eq.~(\ref{A-tilde-via-S-infty})
\begin{equation}
I_{2}^{\text{ren}}=\left\{  \frac{\pi}{2}\frac{1}{S\left(  C\right)  }\left[
\prod_{k=1}^{3}\frac{\Gamma\left(  1-\beta_{k}\right)  }{\Gamma\left(
\beta_{k}\right)  }\right]  \right\}  \left\{  \frac{\pi}{8}\prod_{k=1}
^{3}\left[  \frac{\beta_{k}^{2}\Gamma\left(  \beta_{k}-1\right)  }
{\Gamma\left(  2-\beta_{k}\right)  }\right]  \right\}  .
\end{equation}
Here
\begin{align}
\left[  \prod_{k=1}^{3}\frac{\Gamma\left(  1-\beta_{k}\right)  }{\Gamma\left(
\beta_{k}\right)  }\right]  \prod_{k=1}^{3}\left[  \frac{\beta_{k}^{2}
\Gamma\left(  \beta_{k}-1\right)  }{\Gamma\left(  2-\beta_{k}\right)
}\right]   &  =\prod_{k=1}^{3}\left[  \beta_{k}^{2}\frac{\Gamma\left(
1-\beta_{k}\right)  }{\Gamma\left(  2-\beta_{k}\right)  }\frac{\Gamma\left(
\beta_{k}-1\right)  }{\Gamma\left(  \beta_{k}\right)  }\right] \nonumber\\
&  =\prod_{k=1}^{3}\left[  -\left(  \frac{\beta_{k}}{1-\beta_{k}}\right)
^{2}\right]  .
\end{align}
Thus
\begin{equation}
I_{2}^{\text{ren}}=-\frac{\pi^{2}}{16S\left(  C\right)  }\prod_{k=1}
^{3}\left(  \frac{\beta_{k}}{1-\beta_{k}}\right)  ^{2}.
\label{I-2-ren-triangle-res}
\end{equation}

\subsection{Final result}
\label{final-result-section}

Our final results for $I_{1}^{\text{ren}}$ (\ref{I-1-ren-triangle-res}) and
$I_{2}^{\text{ren}}$ (\ref{I-2-ren-triangle-res}) are symmetric with respect
to permutations of parameters $\beta_{1}$, $\beta_{2}$, $\beta_{3}$ whereas
the intermediate steps of the computation were asymmetric. The symmetry of the
final result is a good test of the consistency of our renormalization
procedure. In fact, it is a test of the compatibility of our analytical
renormalization with SL$\left(  2,\mathbb{R}\right)  $ symmetry discussed
in section~\ref{pro-contra-section}. In our case of triangular contours, the
compatibility of the renormalization procedure with SL$\left(  2,\mathbb{R}
\right)$ symmetry was demonstrated via the explicit computation of $I_{1}^{\text{ren}}$
and $I_{2}^{\text{ren}}$. In principle, one can prove that our analytical
renormalization respects SL$\left(  2,\mathbb{R}\right) $ symmetry for arbitrary
polygons but this is a subject of a separate work devoted to arbitrary polygons.

Now we insert results (\ref{I-1-ren-triangle-res}),
(\ref{I-2-ren-triangle-res}) into the basic expression (\ref{S-ren-via-I-ren})
for $\left\langle \mathcal{S}_{4}\right\rangle ^{\text{ren}}$:
\begin{align}
\left\langle \mathcal{S}_{4}\right\rangle ^{\text{ren}}  &  =-\frac{1}{\sigma
}\frac{D-2}{\left(  8\pi\right)  ^{2}}\left[  I_{1}^{\text{ren}}+\frac{2}
{9}\left(  D-2\right)  I_{2}^{\text{ren}}\right] \nonumber\\
&  =-\frac{1}{\sigma}\frac{D-2}{\left(  8\pi\right)  ^{2}}\left[
\frac{\pi^{2}}{3S\left(  C\right)  }\prod_{k=1}^{3}\frac{\beta_{k}}
{1-\beta_{k}}-\frac{2}{9}\left(  D-2\right)  \frac{\pi^{2}}{16S\left(
C\right)  }\prod_{k=1}^{3}\left(  \frac{\beta_{k}}{1-\beta_{k}}\right)
^{2}\right] \nonumber\\
&  =-\frac{1}{\sigma S\left(  C\right)  }\frac{D-2}{192}\left[  \prod
_{k=1}^{3}\frac{\beta_{k}}{1-\beta_{k}}-\frac{D-2}{24}\left(  \prod_{k=1}
^{3}\frac{\beta_{k}}{1-\beta_{k}}\right)  ^{2}\right]  .
\end{align}
Combining this with eq.~(\ref{f2-C-S-ren}), we complete the derivation of our final result
eq.~(\ref{f2-triangle-re-intro}).

Note that in the case of triangles general polygonal constraint
(\ref{beta-k-range}) is enhanced to
\begin{equation}
0<\beta_{k}<1.
\end{equation}
Obviously $f_{2}\left(  C\right)  $ (\ref{f2-triangle-re-intro}) is regular in this
region of $\beta_{k}$.

\section{Conclusions}

Using EST, we have computed term $f_{2}\left(  C\right) $ of the
large-size expansion (\ref{W-lambda-C-general-expansion-0}) for triangular
Wilson loops. The result is given by expression (\ref{f2-triangle-re-intro}).
The success of the calculation is based on the new version of analytical
regularization using SC\ mapping. This regularization can be applied to
arbitrary polygons.

\section*{Acknowledgments}

I had the luck of starting this work in the stimulating environment formed and guided
by L.~N.~Lipatov and D.~I.~Diakonov up to their last days. I appreciate the support and
the encouragement by V.~Yu.~Petrov.

\appendix                                                       

\renewcommand{\thesection}{\Alph{section}}  \renewcommand{\theequation}
{\Alph{section}.\arabic{equation}} 

\setcounter{section}{0} 

\section{Properties of Laplace determinants}

\label{Laplace-determinant-properties-appendix}

\setcounter{equation}{0} 

\subsection{Heat-kernel expansion for Laplace operator}

\label{heat-kernel-appendix}

For Laplace operator acting in the region bounded by contour $C$ one has the
following heat-kernel expansion \cite{Luscher:1980fr,Pleijel-1954,Kac-1966,SW-1971,BB-1971,Guilkey-1974}:
\begin{equation}
\mathrm{Tr}\,e^{t\Delta\left(  C\right)  }\overset{t\rightarrow0}{=}
\frac{1}{4\pi t}S(C)-\frac{1}{8\sqrt{\pi t}}L(C)-\frac{1}{2}\delta\left(
C\right)  +O\left(  t^{1/2}\right)  . \label{Laplace-heat-kernel}
\end{equation}
For polygonal contours $C$
\begin{equation}
\delta\left(  C\right)  =-\sum\limits_{k=1}^{n_{\text{v}}\left(  C\right)
}\xi(\theta_{k}). \label{Kac-dimension}
\end{equation}
Here $\theta_{k}$ are interior angles of polygon $C$, $n_{\text{v}}\left(
C\right)  $ is the number of vertices and
\begin{equation}
\xi(\theta)=\frac{\pi^{2}-\theta^{2}}{12\pi\theta}. \label{Gamma-cusp-string}
\end{equation}
For function $\xi(\theta)$, an integral representation was derived by M.~Kac
in ref.~\cite{Kac-1966}. Explicit expression (\ref{Gamma-cusp-string}) for
$\xi(\theta)$ was obtained by D. B. Ray (the derivation is described
in ref.~\cite{McKean-Singer-67}).

\subsection{Renormalization of Laplace determinants}

\label{Laplace-determinant-renormalization-appendix}

The determinant of Laplace operator in the proper-time regularization
\begin{equation}
\ln\mathrm{Det}_{\tau}\left[  -\Delta\left(  C\right)  \right]  =-\int_{\tau
}^{\infty}\frac{dt}{t}\mathrm{Tr}\,e^{t\Delta\left(  C\right)  }
\label{ln-det-proper-time}
\end{equation}
has a divergence in the limit $\tau\rightarrow +0$ which is controlled by heat-kernel expansion (\ref{Laplace-heat-kernel}):
\begin{equation}
\ln\mathrm{Det}_{\tau}\left[  -\Delta\left(  C\right)  \right]  \overset
{\tau\rightarrow +0}{=}-\frac{1}{4\pi\tau}S(C)+\frac{1}{4\sqrt{\pi\tau}
}L(C)-\frac{1}{2}\delta\left(  C\right)  \ln\tau+O\left(  \tau^{0}\right)  .
\label{divergences-proper-time}
\end{equation}

\thinspace In $\zeta$-regularization method, one first defines the
regularizing $\zeta$-function at $\mathrm{Re}\,s>1$
\begin{equation}
Z_{C}(s)=\mathrm{Sp}\,\left[  (-\Delta\left(  C\right)  \right]
^{-s}=\frac{1}{\Gamma(s)}\int_{0}^{\infty}dt\,t^{s-1}\mathrm{Tr\,}
e^{t\Delta\left(  C\right)  }. \label{Z-C-def}
\end{equation}
Then one performs the analytical continuation in $s$ and computes the
derivative
\begin{equation}
Z_{C}^{\prime}(s)\equiv\frac{d}{ds}Z_{C}(s).
\end{equation}
Laplace determinant in the $\zeta$-regularization is given by
\begin{equation}
\mathrm{Det}_{\zeta}\left[  -\Delta\left(  C\right)  \right]  =\exp\left[
-Z_{C}^{\prime}(0)\right] \label{det-Z-prime}
\end{equation}
and has the property
\begin{equation}
\mathrm{Det}_{\zeta}\left[  -\Delta\left(  \lambda C\right)  \right]
=\lambda^{\delta\left(  C\right)  }\mathrm{Det}_{\zeta}\left[  -\Delta\left(
C\right)  \right] \label{Laplace-det-zeta-scaling}
\end{equation}
where $\delta\left(  C\right)  $ is given by eq.~(\ref{Kac-dimension}).

\section{1-loop EST corrections}

\label{one-loop-EST-appendix}

\setcounter{equation}{0} 

According to eq.~(\ref{one-loop-terms}) terms
\[
f_{\ln}\left(  C\right)  \ln\lambda+f_{0}\left(  C\right)
\]
of expansion (\ref{W-lambda-C-general-expansion-0}) arise from the 1-loop
EST contribution. These terms are generated by Laplace determinant coming from
Gaussian integral (\ref{W-one-loop-Gaussian}) so that roughly speaking
\begin{equation}
-\frac{D-2}{2}\ln\text{Det}\,\left[  -\Delta\left(  \lambda C\right)  \right]
\rightarrow f_{\ln}\left(  C\right)  \ln\lambda+f_{0}\left(  C\right)  .
\label{ln-Det-f-ln-naive}
\end{equation}
However, precise expressions are sensitive to certain technical subtleties
because both MGT and EST have ultraviolet divergences which must be
renormalized in a consistent way matching MGT and EST. The explicit expression
for $f_{\ln}\left(  C\right)  $ can be extracted from the naive result
(\ref{ln-Det-f-ln-naive}) by comparing eq.~(\ref{ln-Det-f-ln-naive}) with
eq.~(\ref{Laplace-det-zeta-scaling}):
\begin{equation}
f_{\ln}\left(  C\right)  =-\frac{D-2}{2}\delta\left(  C\right)  =\frac{D-2}
{2}\sum\limits_{k=1}^{n_{\text{v}}\left(  C\right)  }\xi(\theta_{k})\,
\label{f-ln}
\end{equation}
where $\xi\left(  \theta\right)  $ given by eq.~(\ref{Gamma-cusp-string}).

Now let us turn to $f_{0}\left(  C\right)  $. Wilson loop $W\left(  C\right)
$ is not a renormalization invariant quantity so that the single functional
$f_{0}\left(  C\right)  $ is not a physical quantity and any attempt to write
an explicit expression for single functional $f_{0}\left(  C\right)  $ will
result in an ugly combination of scheme dependent parameters. However, simple
elegant formulas can be written for linear combinations
(\ref{m-f-C-combination}) associated with renormalization invariant
combinations of Wilson loops --- see eqs.~(\ref{m-f-0-sum-via-Det}) and
(\ref{m-f0-C-sum-res}).

The simplest examples of renormalization invariant ratios are combination
(\ref{W-ratio-1}) and Creutz ratios
\begin{equation}
\frac{W\left(  L_{1},L_{2}\right)  W\left(  L_{1}+L_{1}^{\prime},L_{2}
+L_{2}^{\prime}\right)  }{W\left(  L_{1}+L_{1}^{\prime},L_{2}\right)  W\left(
L_{1},L_{2}+L_{2}^{\prime}\right)  }
\end{equation}
made of Wilson loops $W\left(  L,L^{\prime}\right)  $ for rectangles with
sides $L,L^{\prime}$.

The general form of renormalization invariant ratios in the case of polygonal
contours $C_{a}$ is
\begin{equation}
\prod_{a=1}^{N}\left[  W\left(  C_{a}\right)  \right]  ^{m_{a}}
\label{W-m-product}
\end{equation}
where parameters $m_{a}$ must obey several conditions. Two constraints have a
simple form:
\begin{equation}
\sum_{a=1}^{N}m_{a}=0, \label{balance-vertices-0}
\end{equation}
\begin{equation}
\sum_{a=1}^{N}m_{a}L\left(  C_{a}\right)  =0. \label{balance-perimeter-0}
\end{equation}
Here $L\left(  C_{a}\right)  $ is the perimeter of contour $C_{a}$.

In addition, one must obey a vertex-balance condition. Roughly speaking, each
vertex angle value must appear in the numerator of (\ref{W-m-product}) as many
times as in the denominator. In order to make a more careful formulation of
this constraint, let us first define $\Theta=\left\{  \theta_{A}\right\}  $ as
a set of all vertex angle values appearing in polygons $C_{a}$ so that each
$\theta_{A}$ is included in $\Theta$ only once (whatever often it may appear
in polygons $C_{a}$):
\begin{equation}
\theta_{A}\neq\theta_{B}\,\text{if }A\neq B.
\end{equation}
Next, let $p_{aA}$ be the number of occurrences of angle value $\theta_{A}$
among vertices of polygon $C_{a}$ (so that possible values are $p_{aA}
=0,1,2,\ldots$). Then the vertex-balance constraint reads
\begin{equation}
\sum_{a=1}^{N}m_{a}p_{aA}=0.\label{balance-angles-0}
\end{equation}

A practically important equivalent form of vertex-balance constraint
(\ref{balance-angles-0}) is
\begin{equation}
\sum_{a=1}^{N}\sum_{i=1}^{n_{\text{v}}\left(  C_{a}\right)  }m_{a}\phi\left(
\theta_{ai}\right)  =0 \label{balance-angles-alt}
\end{equation}
where $\theta_{ai}$ with $i=1,2,3,\ldots,n_{\text{v}}\left(  C_a\right)  $ are
interior angles of polygon $C_{a}$ and $\phi\left(  \theta\right)  $ is an
arbitrary function of angle $\theta$.

Sometimes one also needs area-balance condition
\begin{equation}
\sum_{a=1}^{N}m_{a}S\left(  C_{a}\right)  =0. \label{balance-area-0}
\end{equation}
Once all balance conditions (\ref{balance-vertices-0}),
(\ref{balance-perimeter-0}), (\ref{balance-angles-0}), (\ref{balance-area-0})
are satisfied, one finds from eqs.~(\ref{W-int-sigma}),
(\ref{int-sigma-one-loop})
\begin{equation}
\prod_{a=1}^{N}\left[  W\left(  \lambda C_{a}\right)  \right]  ^{m_{a}
}\overset{\lambda\rightarrow\infty}{=}\left\{  \prod_{a=1}^{N}\left\{
\mathrm{Det}\left[  -\Delta\left(  C_{a}\right)  \right]  \right\}  ^{m_{a}
}\right\}  ^{-(D-2)/2}\left[  1+O\left(  \lambda^{-2}\right)  \right]  .
\label{W-product-det-product-0}
\end{equation}
Note that conditions (\ref{balance-vertices-0}), (\ref{balance-perimeter-0}),
(\ref{balance-angles-0}), (\ref{balance-area-0}) guarantee that ultraviolet
divergences of MGT cancel on the LHS of eq.~(\ref{W-product-det-product-0})
whereas ultraviolet divergences of EST cancel on the RHS.

More exactly, the LHS of eq.~(\ref{W-product-det-product-0}) is a
renormalization invariant quantity in MGT\ because renormalization factors
associated with perimeter and cusp divergences cancel due to conditions
(\ref{balance-perimeter-0}), (\ref{balance-angles-alt}). On the RHS,
conditions (\ref{balance-vertices-0}), (\ref{balance-perimeter-0}),
(\ref{balance-angles-0}), (\ref{balance-area-0}) guarantee a cancellation of
all divergences and stability of the result with respect to a wide class of
renormalization schemes including the renormalizations based on proper-time
regularization (\ref{ln-det-proper-time}) and $\zeta$-regularization
(\ref{det-Z-prime}).

If one imposes only constraints (\ref{balance-vertices-0}),
(\ref{balance-perimeter-0}), (\ref{balance-angles-0}), but omits
area-balance condition (\ref{balance-area-0}) then eq.~(\ref{W-product-det-product-0}) must be
modified. This modification depends on the renormalization scheme for Laplace
determinants. In the case of $\zeta$-renormalization (\ref{det-Z-prime}) one
has under conditions (\ref{balance-vertices-0}), (\ref{balance-perimeter-0}),
(\ref{balance-angles-0})

\begin{align}
&  \left[  \exp\left(  \lambda^{2}\sigma\sum_{a=1}^{N}m_{a}S\left(
C_{a}\right)  \right)  \right]  \prod_{a=1}^{N}\left[  W\left(  \lambda
C_{a}\right)  \right]  ^{m_{a}}\nonumber\\
&  \overset{\lambda\rightarrow\infty}{=}\left\{  \prod_{a=1}^{N}\left\{
\mathrm{Det}_{\zeta}\left[  -\Delta\left(  C_{a}\right)  \right]  \right\}
^{m_{a}}\right\}  ^{-(D-2)/2}\left[  1+O\left(  \lambda^{-2}\right)  \right]
. \label{W-product-det-product-1}
\end{align}
Here $\sigma$ is the physical string tension of MGT (as it appears in area law
(\ref{Wilson-area-law})) and $S\left(  C_{a}\right)  $ is area of polygon
$C_{a}$.

Comparing eq.~(\ref{W-product-det-product-1}) with expansion
(\ref{W-lambda-C-general-expansion-0}) and using balance conditions
(\ref{balance-vertices-0}), (\ref{balance-perimeter-0}), we find
\begin{equation}
\sum_{a=1}^{N}m_{a}f_{0}\left(  C_{a}\right)  =-\frac{D-2}{2}\sum_{a=1}
^{N}m_{a}\ln\mathrm{Det}_{\zeta}\left[  -\Delta\left(  C_{a}\right)  \right]
. \label{m-f-0-sum-via-Det}
\end{equation}

The computation of the RHS\ is discussed
in appendix~\ref{Computation-Laplace-determinants-appendix}.

\section{Calculation of Laplace determinants for polygons}

\label{Computation-Laplace-determinants-appendix}

\setcounter{equation}{0} 

\subsection{2D Laplace determinant for general polygons}

\subsubsection{Results of E. Aurell and P. Salomonson}

\label{Notation-AS-appendix}

It is well-known that Laplace determinants in 2D regions can be computed using
the conformal anomaly \cite{Polyakov-1981,DNOP-1982,DOP-1982,Alvarez-1983}. In order to
compute Laplace determinant $\mathrm{Det}\left(  -\Delta\left(  C\right)
\right)  $ for the region bounded by contour $C$, one has to construct a
conformal mapping of this region to some standard region (semiplane or
circle). In the case of polygons $C$ this conformal mapping is known as
Schwarz-Christoffel (SC)\ mapping. Thus combining the conformal anomaly and SC
mapping, one can compute Laplace determinants for arbitrary polygons. However,
on this way one must solve two problems:

\noindent$\bullet$ the anomaly-based representation for Laplace determinants has an integral
form and this integral must be computed,

\noindent$\bullet$ this integral representation has cusp divergences which must be renormalized.

These two problems were successfully solved by E. Aurell and P. Salomonson
in ref.~\cite{AS-93-journal} where the renormalized functional determinant (in
$\zeta$-regularization) of two-dimensional Laplace operator with Dirichlet
boundary condition for an arbitrary polygonal region was computed and expressed
via parameters of SC mapping.

In this appendix we compute the combinations of Laplace determinants appearing
on the RHS of eqs.~(\ref{W-product-det-product-1}), (\ref{m-f-0-sum-via-Det})
\begin{equation}
\prod_{a=1}^{N}\left\{  \mathrm{Det}_{\zeta}\left[  -\Delta\left(
C_{a}\right)  \right]  \right\}  ^{m_{a}},
\end{equation}
using the expressions of ref.~\cite{AS-93-journal} for $\mathrm{Det}_{\zeta
}\left[  -\Delta\left(  C_{a}\right)  \right]  $ and making simplifications
based on vertex-balance condition (\ref{balance-angles-0}).

In order to simplify the extraction of final expressions for Laplace
determinants from ref.~\cite{AS-93-journal}, we provide a small dictionary
relating the notation used in the current work (LHS) and the notation adopted
in ref.~\cite{AS-93-journal} (RHS with label AS).

Polygonal boundary $C$ of the region where Laplace operator acts:
\begin{equation}
C=P_{\text{AS}}.
\end{equation}

Area of the region bounded by contour $C$:
\begin{equation}
S\left(  C\right)  =A_{\text{AS}}.
\end{equation}

Complex coordinate in the polygon plane:
\begin{equation}
\zeta=z_{\text{AS}}.
\end{equation}

Logarithm of Laplace determinant in $\zeta$-renormalization (\ref{det-Z-prime}):
\begin{equation}
\ln\mathrm{Det}_{\zeta}\left[  -\Delta\left(  C\right)  \right]  =-\left[
Z_{P}^{\prime}\left(  0\right)  \right]  _{\text{AS}}. \label{Z-prime-P-AS}
\end{equation}

Function $\delta\left(  C\right)  $ given by eqs.~(\ref{Kac-dimension}),
(\ref{Gamma-cusp-string}):
\begin{equation}
\delta\left(  C\right)  =-2\left[  Z_{P}\left(  0\right)  \right]
_{\text{AS}}.
\end{equation}

Function $Z_{1-\beta}^{\prime}(0)$ (do not confuse it with quantity
$Z_{P}^{\prime}\left(  0\right)  $ appearing in eq.~(\ref{Z-prime-P-AS})) playing
an important role in ref.~\cite{AS-93-journal} but irrelevant for our work:
\begin{equation}
h\left(  \beta\right)  =\left[  Z_{1-\beta}^{\prime}(0)\right]  _{\text{AS}}.
\label{h-Z-prime-dic}
\end{equation}

For angles of polygons ref.~\cite{AS-93-journal} uses the same parameters
$\theta_{k},\beta_{k},\alpha_{k}$ as in this work (but with Greek indices) ---
see eqs.~(\ref{theta-k-via-beta-k}), (\ref{alpha-k-via-beta-k-theta-k}).

\subsubsection{SC mapping from unit circle to polygon}

One should keep in mind that ref.~\cite{AS-93-journal} uses
Schwarz-Christoffel (SC) mapping of the unit circle $\left|  u\right|  \leq1$
in complex $u$-plane to the polygon in complex $\zeta$-plane
\begin{equation}
\frac{d\zeta}{du}=e^{\lambda_{0}}\prod_{k=1}^{n_{\text{v}}}\left(
u-e^{i\phi_{k}}\right)  ^{-\beta_{k}} \label{CS-circular}
\end{equation}
(the original notation of ref.~\cite{AS-93-journal} uses $z$ instead of
$\zeta$). Here $e^{i\phi_{k}}$ are points on the boundary of the unit circle
which are mapped to vertices $\zeta_{k}$ of polygon $C$. $\lambda_{0}$ is a
real constant controlling the size of the polygon.

\subsubsection{Result for Laplace determinant}

The general result for the determinant of Laplace operator defined in polygon
$C$ with Dirichlet boundary condition in $\zeta$-regularization
(\ref{Z-C-def}) -- (\ref{det-Z-prime}) is given by eq.~(62) of ref.~\cite{AS-93-journal}:
\begin{align}
-\ln\mathrm{Det}_{\zeta}\left[  -\Delta\left(  C\right)  \right]   &
=\sum\limits_{j=1}^{n_{\text{v}}}h\left(  \beta_{j}\right)  +\frac{\lambda
_{0}}{12}\sum\limits_{j=1}^{n_{\text{v}}}\frac{\left(  2-\beta_{j}\right)
\beta_{j}}{1-\beta_{j}}\nonumber\\
&  -\frac{1}{12}\sum\limits_{1\leq j\neq k\leq n_{\text{v}}}\frac{\beta
_{j}\beta_{k}}{1-\beta_{j}}\ln\left|  e^{i\phi_{j}}-e^{i\phi_{k}}\right|  .
\label{AS-det-org}
\end{align}
The RHS contains already defined circular SC\ parameters $\lambda_{0}$,
$\left\{  \phi_{k}\right\}$, $\left\{  \beta_{k}\right\} $ and function
$h\left(  \beta\right)  $ (\ref{h-Z-prime-dic}) whose explicit form plays no
role for our work.

\subsubsection{SC\ mapping: from unit circle to semiplane}

\label{SC-mapping-semiplane-section}

One can easily establish a connection between semiplane version of SC\ mapping
(\ref{SC-mapping-general}) and unit-circle version of SC mapping
(\ref{CS-circular}). Indeed, one can map the unit circle in the $u$-plane to the
upper semiplane of complex variable $z$ using linear-fractional
transformation
\begin{equation}
z=i\frac{1-u}{1+u} \label{omega-via-u}
\end{equation}
with the inverse mapping
\begin{equation}
u=\frac{1+iz}{1-iz}. \label{u-via-omega}
\end{equation}
Transformation (\ref{omega-via-u})
maps points $e^{i\phi_{k}}$ on the boundary of the unit circle in the
$u$-plane to points $z_{k}$ on the real axis of the $z$-plane
\begin{equation}
z_{k}=\tan\left(  \frac{1}{2}\phi_{k}\right)  , \label{omega-via-phi}
\end{equation}
\begin{equation}
e^{i\phi_{k}}=\frac{1+iz_{k}}{1-iz_{k}}. \label{exp-i-phi-via-omega}
\end{equation}

Defining
\begin{equation}
A=\frac{1}{2}e^{\lambda_{0}}\left[  -i\exp\left(  \frac{i}{2}\sum
\limits_{j=1}^{n_{\text{v}}}\beta_{j}\phi_{j}\right)  \right]  \left[
\prod_{j=1}^{n_{\text{v}}}\left[  \cos\left(  \frac{1}{2}\phi_{j}\right)
\right]  ^{-\beta_{j}}\right]  , \label{A-via-0-phi}
\end{equation}
one finds that in terms of the $z$-semiplane parametrization, SC mapping
(\ref{CS-circular}) takes form (\ref{SC-mapping-general}).

\subsubsection{Laplace determinants in terms of semiplane parameters of SC\ mapping}

\label{Lapalce-det-via-SC-semiplane-subsection}

Using relation (\ref{omega-via-phi}), we find
\begin{equation}
\ln\left|  e^{i\phi_{j}}-e^{i\phi_{k}}\right|  =\ln\left|  z_{k}-z_{j}\right|
+\ln\left|  \cos\left(  \frac{1}{2}\phi_{j}\right)  \right|  +\ln\left|
\cos\left(  \frac{1}{2}\phi_{k}\right)  \right|  +\ln
2.\label{phi-k-z-k-sum-ln-relation}
\end{equation}
It follows from eq.~(\ref{A-via-0-phi}) that
\begin{equation}
\lambda_{0}=\ln\left|  A\right|  +\ln2+\sum_{j=1}^{n_{\text{v}}}\beta_{j}
\ln\left|  \cos\left(  \frac{1}{2}\phi_{j}\right)  \right|  .
\end{equation}
Combining the last two equations and using constraint (\ref{beta-sum-rule}),
one can derive relation
\begin{align}
&  \lambda_{0}\sum\limits_{j=1}^{n_{\text{v}}}\frac{\left(  2-\beta
_{j}\right)  \beta_{j}}{1-\beta_{j}}-\sum\limits_{1\leq j\neq k\leq
n_{\text{v}}}\frac{\beta_{j}\beta_{k}}{1-\beta_{j}}\ln\left|  e^{i\phi_{j}
}-e^{i\phi_{k}}\right|  \nonumber\\
&  =-\sum\limits_{1\leq j\neq k\leq n_{\text{v}}}\frac{\beta_{j}\beta_{k}
}{1-\beta_{j}}\ln\left|  \frac{z_{j}-z_{k}}{A}\right|
.\label{identity-for-log-det}
\end{align}
Now eq.~(\ref{AS-det-org}) takes the form
\begin{equation}
\ln\mathrm{Det}_{\zeta}\left[  -\Delta\left(  C\right)  \right]
=-\sum\limits_{j=1}^{n_{\text{v}}}h\left(  \beta_{j}\right)  +\frac{1}{12}
\sum\limits_{1\leq j\neq k\leq n_{\text{v}}}\frac{\beta_{j}\beta_{k}}
{1-\beta_{j}}\ln\left|  \frac{z_{j}-z_{k}}{A}\right|
.\label{Z-P-prime-semiplane-0}
\end{equation}
This provides an expression for Laplace determinant via parameters
$z_{k}$, $\beta_{k}$, $A$ of SC\ mapping in semiplane form (\ref{SC-mapping-general}
). The RHS of (\ref{Z-P-prime-semiplane-0})\ contains function $h\left(
\beta\right)  $ (\ref{h-Z-prime-dic}) which is computed
in ref.~\cite{AS-93-journal} but cancels in our final formulas.

\subsubsection{Taking one SC\ vertex to infinity}

Sometimes it is convenient to keep one SC\ vertex at infinity and to work with
SC\ representation (\ref{SC-tilded-1}). Taking the limit $\left|
z_{n_{\text{v}}}\right|  \rightarrow\infty$ on the RHS eq.
(\ref{Z-P-prime-semiplane-0}) with constraint (\ref{A-tilde-A-condition}) and
using relation (\ref{beta-n-v-via-other}), we find the expression for Laplace
determinant in terms of SC\ mapping (\ref{SC-tilded-1})

\begin{align}
\ln\mathrm{Det}_{\zeta}\left[  -\Delta\left(  C\right)  \right]   &
=-\sum\limits_{j=1}^{n_{\text{v}}}h\left(  \beta_{j}\right)  +\frac{1}
{12}\left[  \sum\limits_{1\leq j\neq k\leq n_{\text{v}}-1}\frac{\beta_{j}
\beta_{k}}{1-\beta_{j}}\ln\left|  z_{j}-z_{k}\right|  \right. \nonumber\\
&  \left.  \,-\left(  2+\sum_{j=1}^{n_{\text{v}}}\frac{\beta_{j}}{1-\beta_{j}
}\right)  \ln\left|  \tilde{A}\right|  \right]  .
\label{log-det-res-SC-z-n-at-infinity}
\end{align}

\subsubsection{Balanced combinations of Laplace determinants}

Let us split

\begin{equation}
\ln\mathrm{Det}_{\zeta}\left[  -\Delta\left(  C\right)  \right]  =\left\{
\ln\mathrm{Det}_{\zeta}\left[  -\Delta\left(  C\right)  \right]  \right\}
^{\left(  1\right)  }+\left\{  \ln\mathrm{Det}_{\zeta}\left[  -\Delta\left(
C\right)  \right]  \right\}  ^{\left(  2\right)  }
\label{Z-P-prime-semiplane-again}
\end{equation}
where
\begin{equation}
\left\{  \ln\mathrm{Det}_{\zeta}\left[  -\Delta\left(  C\right)  \right]
\right\}  ^{\left(  2\right)  }=-\sum\limits_{j=1}^{n_{\text{v}}}h\left(
\beta_{j}\right)  . \label{log-det-AS-part-2}
\end{equation}
Then in the case of SC\ mapping (\ref{SC-mapping-general}) with finite
SC\ vertices we have according to eq.~(\ref{Z-P-prime-semiplane-0})
\begin{equation}
\left\{  \ln\mathrm{Det}_{\zeta}\left[  -\Delta\left(  C\right)  \right]
\right\}  ^{\left(  1\right)  }=\frac{1}{12}\sum\limits_{1\leq j\neq k\leq
n_{\text{v}}}\frac{\beta_{j}\beta_{k}}{1-\beta_{j}}\ln\left|  \frac{z_{j}
-z_{k}}{A}\right|  . \label{log-det-AS-part-1}
\end{equation}
In the case of SC\ mapping (\ref{SC-tilded-1}) with vertex $z_{n_{\text{v}}}$
at infinity we derive from eq.~(\ref{log-det-res-SC-z-n-at-infinity})
\begin{align}
&  \left\{  \ln\mathrm{Det}_{\zeta}\left[  -\Delta\left(  C\right)  \right]
\right\}  ^{\left(  1\right)  }\nonumber\\
&  =\frac{1}{12}\left[  \sum\limits_{1\leq j\neq k\leq n_{\text{v}}
-1}\frac{\beta_{j}\beta_{k}}{1-\beta_{j}}\ln\left|  z_{j}-z_{k}\right|
\,-\left(  2+\sum_{j=1}^{n_{\text{v}}}\frac{\beta_{j}}{1-\beta_{j}}\right)
\ln\left|  \tilde{A}\right|  \right]  . \label{log-det-AS-part-1-z-n-infinity}
\end{align}

Now we can compute the combination of Laplace determinants appearing on the
RHS\ of eq.~(\ref{m-f-0-sum-via-Det}). For each polygon $C_{a}$ appearing in
eq.~(\ref{m-f-0-sum-via-Det}) we introduce SC\ mapping
(\ref{SC-mapping-general})

\begin{equation}
\frac{d\zeta}{dz}=A_{a}\prod_{k=1}^{n_{\text{v}}\left(  C_{a}\right)  }\left(
z-z_{ak}\right)  ^{-\beta_{ak}}. \label{C-a-SC}
\end{equation}
Using decomposition (\ref{Z-P-prime-semiplane-again}), we obtain
\begin{align}
\ln\prod_{a=1}^{N}\left\{  \mathrm{Det}_{\zeta}\left[  -\Delta\left(
C_{a}\right)  \right]  \right\}  ^{m_{a}}  &  =\sum_{a=1}^{N}m_{a}\left\{
\ln\mathrm{Det}_{\zeta}\left[  -\Delta\left(  C_{a}\right)  \right]  \right\}
^{\left(  1\right)  }\nonumber\\
&  +\sum_{a=1}^{N}m_{a}\left\{  \ln\mathrm{Det}_{\zeta}\left[  -\Delta\left(
C_{a}\right)  \right]  \right\}  ^{\left(  2\right)  }.
\end{align}
Here
\begin{equation}
\sum_{a=1}^{N}m_{a}\left\{  \ln\mathrm{Det}_{\zeta}\left[  -\Delta\left(
C_{a}\right)  \right]  \right\}  ^{\left(  2\right)  }=\sum_{a=1}^{N}m_{a}
\sum\limits_{j=1}^{n_{\text{v}}\left(  C_{a}\right)  }h\left(  \beta
_{aj}\right)  =0
\end{equation}
vanishes due to vertex-balance condition (\ref{balance-angles-alt}).

Hence
\begin{equation}
\ln\prod_{a=1}^{N}\left\{  \mathrm{Det}_{\zeta}\left[  -\Delta\left(
C_{a}\right)  \right]  \right\}  ^{m_{a}}=\sum_{a=1}^{N}m_{a}\left\{
\ln\mathrm{Det}_{\zeta}\left[  -\Delta\left(  C_{a}\right)  \right]  \right\}
^{\left(  1\right)  }. \label{ln-det-reduction-no-h}
\end{equation}
On the RHS we can use any of two SC representations (\ref{log-det-AS-part-1}),
(\ref{log-det-AS-part-1-z-n-infinity}) for Laplace determinants. In the case
when all SC vertices are finite for all polygons we find inserting
(\ref{log-det-AS-part-1}) into eq.~(\ref{ln-det-reduction-no-h})

\begin{equation}
\ln\prod_{a=1}^{N}\left\{  \mathrm{Det}_{\zeta}\left[  -\Delta\left(
C_{a}\right)  \right]  \right\}  ^{m_{a}}=\frac{1}{12}\sum_{a=1}^{N}m_{a}
\sum\limits_{1\leq j\neq k\leq n_{\text{v}}\left(  C_{a}\right)  }
\frac{\beta_{aj}\beta_{ak}}{1-\beta_{aj}}\ln\left|  \frac{z_{aj}-z_{ak}}
{A_{a}}\right|  . \label{m-log-det-sum-res}
\end{equation}
Remember that this result was derived assuming vertex-balance condition
(\ref{balance-angles-0}).

\subsubsection{Result for balanced linear combinations of $f_{0}\left(
C_{a}\right)  $}

Inserting eq.~(\ref{m-log-det-sum-res}) into eq.~(\ref{m-f-0-sum-via-Det}), we
find
\begin{equation}
\sum_{a=1}^{N}m_{a}f_{0}\left(  C_{a}\right)  =-\frac{D-2}{24}\sum_{a=1}
^{N}m_{a}\sum\limits_{1\leq j\neq k\leq n_{\text{v}}\left(  C_{a}\right)
}\frac{\beta_{aj}\beta_{ak}}{1-\beta_{aj}}\ln\left|  \frac{z_{aj}-z_{ak}
}{A_{a}}\right|  . \label{m-f0-C-sum-res}
\end{equation}
This result is derived for the set of polygons $C_{a}$ described by
SC\ mappings (\ref{C-a-SC}), assuming balance conditions
(\ref{balance-vertices-0}), (\ref{balance-perimeter-0}),
(\ref{balance-angles-0}).

\subsection{Case of triangles}

\label{triangle-appendix}

\subsubsection{Result}

The primary subject of this paper is the computation of 2-loop correction
$f_{2}\left(  C\right)  $ for triangles. For completeness it makes sense to
compute also 1-loop terms $f_{\ln}\left(  C\right)  $ and $f_{0}\left(
C\right)  $ for triangles. Quantity $f_{\ln}\left(  C\right)  $ is given by
simple formula (\ref{f-ln}) which can be used for any polygon including
triangle. As for $f_{0}\left(  C\right)  $, we must compute the RHS of eq.
(\ref{m-log-det-sum-res}) assuming balance conditions. If all polygons $C_{a}$
are triangles and vertex-balance condition (\ref{balance-angles-0}) holds then one can derive relation
\begin{align}
&  \ln\prod_{a=1}^{N}\left\{  \mathrm{Det}_{\zeta}\left[  -\Delta\left(
C_{a}\right)  \right]  \right\}  ^{m_{a}}\nonumber\\
&  =\frac{1}{24}\sum_{a=1}^{N}\left[  \sum\limits_{i=1}^{3}\left(  \alpha
_{ai}^{-1}-\alpha_{ai}\right)  \right]  \left[  -\ln S\left(  C_{a}\right)
+\sum\limits_{k=1}^{3}\ln\frac{\Gamma\left(  \alpha_{ak}\right)  }
{\Gamma\left(  1-\alpha_{ak}\right)  }\right]
\label{log-det-sum-triangles-res}
\end{align}
where in agreement with eq.~(\ref{alpha-k-via-beta-k-theta-k})
\begin{equation}
\alpha_{ai}=\frac{1}{\pi}\theta_{ai}
\end{equation}
and $\theta_{ai}$ are interior angles ($i=1,2,3$) of triangle $C_{a}$
($a=1,2,\ldots N$). $\Gamma$ is Euler Gamma function. Anomalous dimensional
structure of logarithmic term $\ln S\left(  C_{a}\right)  $ containing area
$S\left(  C_{a}\right)  $ of polygon $C_{a}$ on the RHS of eq.
(\ref{log-det-sum-triangles-res}) makes no problem due to vertex-balance
condition (\ref{balance-angles-alt}). Eq.~(\ref{log-det-sum-triangles-res}) is
derived in section~\ref{det-triangle-derivation-section}.

Inserting eq.~(\ref{log-det-sum-triangles-res}) into eq.~(\ref{m-f-0-sum-via-Det}
), we find for triangles $C_{a}$ obeying balance conditions
(\ref{balance-vertices-0}), (\ref{balance-perimeter-0}),
(\ref{balance-angles-0})
\[
\sum_{a=1}^{N}m_{a}f_{0}\left(  C_{a}\right)  =-\frac{D-2}{48}
\]
\begin{equation}
\times\sum_{a=1}^{N}\left[  \sum\limits_{i=1}^{3}\left(  \alpha_{ai}
^{-1}-\alpha_{ai}\right)  \right]  \left[  -\ln S\left(  C_{a}\right)
+\sum\limits_{k=1}^{3}\ln\frac{\Gamma\left(  \alpha_{ak}\right)  }
{\Gamma\left(  1-\alpha_{ak}\right)  }\right]  .
\label{m-f0-C-sum-triangles-res}
\end{equation}

\subsubsection{Nontriviality of vertex-balance condition for triangles}

Before deriving relation (\ref{log-det-sum-triangles-res}) it makes sense to
check whether this relation is of any use. The problem is that in the case of
triangles the combination of vertex-balance condition (\ref{balance-angles-0})
and angle sum rules (\ref{angle-sum-rule}) for each triangle is very
restrictive. One can easily obey balance conditions using simplest
renormalization invariant combination (\ref{W-ratio-1}) but the for the
computation of large-size expansion (\ref{W-ratio-2}) for this combination it
is sufficient to know function $f_{\ln}\left(  C\right)  $ whereas
$f_{0}\left(  C\right)  $ does not appear in (\ref{W-ratio-2}) at all. In
other words, for combination (\ref{W-ratio-1}) one has
\begin{equation}
\ln\prod_{a=1}^{N}\left\{  \mathrm{Det}_{\zeta}\left[  -\Delta\left(
C_{a}\right)  \right]  \right\}  ^{m_{a}}=0,
\end{equation}
\begin{equation}
\sum_{a=1}^{N}m_{a}f_{0}\left(  C_{a}\right)  =0.
\end{equation}

One can wonder whether using only triangle contours $C_{a}$ we can satisfy
balance condition in a non-trivial way so that
\begin{equation}
\sum_{a=1}^{N}m_{a}f_{0}\left(  C_{a}\right)  \neq0
\end{equation}
i.e. without using similar triangles like in eq.~(\ref{W-ratio-1}). The answer
to this question is positive but the examples are rather exotic because
vertex-balance constraint (\ref{balance-angles-0}) must be combined with the
angle sum constraint (\ref{angle-sum-rule})
\begin{equation}
\theta_{a1}+\theta_{a2}+\theta_{a3}=\pi
\end{equation}
for each triangle $C_{a}$.

An example of the solution of this problem can be constructed using Wilson
loops $w\left(  n_{a1},n_{a2},n_{a3}\right)  $ for triangles $C_{a}$ with
interior angles
\begin{equation}
\theta_{ak}=\frac{\pi}{9}n_{ak}
\end{equation}
with angle sum rule
\begin{equation}
n_{a1}+n_{a2}+n_{a3}=9. \label{n-angle-sum-rule}
\end{equation}
One can easily check that combination of Wilson loops
\begin{equation}
\frac{w\left(  1,3,5\right)  \left[  w\left(  2,3,4\right)  \right]  ^{2}
}{w\left(  1,4,4\right)  w\left(  2,2,5\right)  w\left(  3,3,3\right)  }
\end{equation}
obeys both vertex-balance condition (\ref{balance-angles-0}) and angle sum
rule (\ref{n-angle-sum-rule}) for each triangle.

\subsubsection{Derivation}

\label{det-triangle-derivation-section}

Now we turn to the derivation of eq.~(\ref{log-det-sum-triangles-res}). In
principle, one can derive eq.~(\ref{log-det-sum-triangles-res}) from general
equation (\ref{m-log-det-sum-res}). On the other hand, one can profit from the
explicit expression computed in ref.~\cite{AS-93-journal} for Laplace
determinant with triangle contour $C$. One can read this explicit expression
from eqs.~(68) and (69) of ref.~\cite{AS-93-journal}. In the original notation
of ref.~\cite{AS-93-journal} the result reads
\begin{equation}
Z_{T}^{\prime}\left(  0\right)  =\sum_{i=1}^{3}Z_{\alpha_{i}}^{\prime
}(0)+Z_{T}\left(  0\right)  \ln\left[  \frac{\pi}{2A}\prod_{k=1}
^{3}\frac{\Gamma\left(  \alpha_{k}\right)  }{\Gamma\left(  1-\alpha
_{k}\right)  }\right]  \label{det-trangle-from-AS}
\end{equation}
where parameters $\alpha_{k}$ are defined by eq.
(\ref{alpha-k-via-beta-k-theta-k}) and label $T$ stands for triangle. Using
relations of appendix~\ref{Notation-AS-appendix} (with $P=T$), we can
translate eq.~(\ref{det-trangle-from-AS}) to the notation adopted in the
current work:
\begin{equation}
\mathrm{Det}_{\zeta}\left[  -\Delta\left(  C\right)  \right]  =\left[
\frac{\pi}{2S\left(  C\right)  }\prod_{k=1}^{3}\frac{\Gamma\left(  \alpha
_{k}\right)  }{\Gamma\left(  1-\alpha_{k}\right)  }\right]  ^{-\frac{1}{2}
\xi\left(  C\right)  }\exp\left[  -\sum_{i=1}^{3}h\left(  1-\alpha_{k}\right)
\right]  .
\end{equation}

Comparing this with eqs.~(\ref{Z-P-prime-semiplane-again}),
(\ref{log-det-AS-part-2}) and using eqs.~(\ref{Kac-dimension}),
(\ref{Gamma-cusp-string}), we get rid of irrelevant functions $h\left(
1-\alpha_{k}\right)  $
\begin{equation}
\left\{  \ln\mathrm{Det}_{\zeta}\left[  -\Delta\left(  C\right)  \right]
\right\}  ^{\left(  1\right)  }=\frac{1}{24}\left[  \sum\limits_{i=1}
^{3}\left(  \alpha_{i}^{-1}-\alpha_{i}\right)  \right]  \ln\left[  \frac{\pi
}{2S\left(  C\right)  }\prod_{k=1}^{3}\frac{\Gamma\left(  \alpha_{k}\right)
}{\Gamma\left(  1-\alpha_{k}\right)  }\right]  .
\label{log-det-triangle-1-res}
\end{equation}
Alternatively one can derive eq.~(\ref{log-det-triangle-1-res}) by applying
general eq.~(\ref{log-det-AS-part-1-z-n-infinity}) to triangular case
$n_{\text{v}}=3$, choosing for simplicity $z_{1}=0$, $z_{2}=1$ and using
relation (\ref{A-tilde-via-S-infty}) between $\tilde{A}$ and $S\left(
C\right)  $.

Next we insert eq.~(\ref{log-det-triangle-1-res}) into eq.
(\ref{ln-det-reduction-no-h}) and get rid of factors $\pi/2$ using
vertex-balance condition (\ref{balance-angles-alt}):
\begin{align}
&  \sum_{a=1}^{N}m_{a}\left\{  \ln\mathrm{Det}_{\zeta}\left[  -\Delta\left(
C_{a}\right)  \right]  \right\}  =\sum_{a=1}^{N}m_{a}\left\{  \ln
\mathrm{Det}_{\zeta}\left[  -\Delta\left(  C_{a}\right)  \right]  \right\}
^{\left(  1\right)  }\nonumber\\
&  =\frac{1}{24}\sum_{a=1}^{N}\left[  \sum\limits_{i=1}^{3}\left(  \alpha
_{ai}^{-1}-\alpha_{ai}\right)  \right]  \left[  -\ln S\left(  C_{a}\right)
+\sum\limits_{k=1}^{3}\ln\frac{\Gamma\left(  \alpha_{ak}\right)  }
{\Gamma\left(  1-\alpha_{ak}\right)  }\right]  .
\end{align}
This completes the derivation of eq.~(\ref{log-det-sum-triangles-res}).

\section{Schwarz derivative}

\setcounter{equation}{0}  \label{Appendix-Schwarz-derivative}

Schwarz derivative of function $z\left(  w\right)  $ is defined by expression

\begin{equation}
\left\{  z,w\right\}  =\frac{z^{\prime\prime\prime}}{z^{\prime}}-\frac{3}
{2}\left(  \frac{z^{\prime\prime}}{z^{\prime}}\right)  ^{2}
\label{Schwarz-derivative-def-3}
\end{equation}
where the prime stands for derivative $d/dw$:
\begin{equation}
z^{\prime}=\frac{dz}{dw},\quad z^{\prime\prime}=\frac{d^{2}z}{dw^{2}},\quad
z^{\prime\prime\prime}=\frac{d^{3}z}{dw^{3}}.
\end{equation}
Schwarz derivative has the property
\begin{equation}
\left\{  z_{1},z_{2}\right\}  =-\left(  \frac{dz_{1}}{dz_{2}}\right)
^{2}\left\{  z_{2},z_{1}\right\}  . \label{theorem-1-statement}
\end{equation}

\section{Propagator in diagonal limit}

\setcounter{equation}{0} 

\label{Appendix-Diagonal-limit}

\subsection{Green function}

It is well-known that Green function of the two-dimensional Laplace operator
in simply connected region $U$ with Dirichlet boundary conditions can be
expressed via the conformal mapping of the upper complex semiplane to region
$U$. We denote this Laplace operator $\Delta\left(  C\right)  $ where $C$ is
the boundary of $U$. Let
\begin{equation}
x=\tilde{f}\left(  w\right)
\label{f-tilde-maping}
\end{equation}
be a representation of this conformal mapping in terms of real coordinates
$w^{1},w^{2}$ on the semiplane and real coordinates $x^{1},x^{2}$ in region
$U$. Then Green function $\left\langle x^{\prime}\right|  \left[
\Delta\left(  C\right)  \right]  ^{-1}\left|  x\right\rangle $ for region $U$
and Green function for the semiplane $\left\langle w^{\prime}\right|
\Delta_{\mathrm{semiplane}}^{-1}\left|  w\right\rangle $ are connected by
relation
\begin{equation}
\left\langle x^{\prime}\right|  \left[  \Delta\left(  C\right)  \right]
^{-1}\left|  x\right\rangle =\left\langle w^{\prime}\right|  \Delta
_{\mathrm{semiplane}}^{-1}\left|  w\right\rangle .
\label{GF-polygon-GF-semiplane-2}
\end{equation}

Green function for the semiplane can be constructed by the well-known image
method from the Green function on the plane:
\begin{equation}
\left\langle w^{\prime}\right|  \Delta_{\mathrm{semiplane}}^{-1}\left|
w\right\rangle =\frac{1}{2\pi}\left[  \ln\left|  w-w^{\prime}\right|
-\ln\left|  w^{R}-w^{\prime}\right|  \right]  .
\end{equation}
Here $w^{R}$ is the reflection of point $w$ with respect to $w^{1}$-axis:
\begin{align}
\left(  w^{R}\right)  ^{1}  &  =w^{1},\\
\left(  w^{R}\right)  ^{2}  &  =-w^{2}.
\end{align}
Hence
\begin{equation}
\left\langle x^{\prime}\right|  \left[  \Delta\left(  C\right)  \right]
^{-1}\left|  x\right\rangle =\frac{1}{2\pi}\left[  \ln\left|  w-w^{\prime
}\right|  -\ln\left|  w^{R}-w^{\prime}\right|  \right]  .
\end{equation}

This result can be rewritten using complex coordinates
\begin{equation}
z=w^{1}+iw^{2}
\end{equation}
on the upper semiplane of complex $z$ and
\begin{equation}
\zeta=x^{1}+ix^{2}
\end{equation}
for region $U$. Then
\begin{equation}
\left\langle \zeta^{\prime}\right|  \left[  \Delta\left(  C\right)  \right]
^{-1}\left|  \zeta\right\rangle =\frac{1}{4\pi}\ln\frac{\left(  z^{\prime
}-z\right)  \left(  z^{\prime\ast}-z^{\ast}\right)  }{\left(  z^{\prime
}-z^{\ast}\right)  \left(  z^{\prime\ast}-z\right)  }
\label{GF-polygon-GF-semiplane-4}
\end{equation}
where
\begin{equation}
\zeta=f\left(  z\right)  ,
\end{equation}
\begin{equation}
\zeta^{\prime}=f\left(  z^{\prime}\right)  ,
\end{equation}
and $f$ is the conformal mapping from the semiplane to\ $U$ corresponding to
mapping $\tilde{f}$ (\ref{f-tilde-maping}) formulated in terms of points with real coordinates.

Let
\begin{equation}
z=\phi\left(  \zeta\right)
\end{equation}
be the inverse mapping. Then eq.~(\ref{GF-polygon-GF-semiplane-4}) takes the
form
\begin{equation}
\left\langle \zeta^{\prime}\right|  \left[  \Delta\left(  C\right)  \right]
^{-1}\left|  \zeta\right\rangle =\frac{1}{4\pi}\ln\frac{\left[  \phi\left(
\zeta^{\prime}\right)  -\phi\left(  \zeta\right)  \right]  \left[  \phi\left(
\zeta^{\prime}\right)  -\phi\left(  \zeta\right)  \right]  ^{\ast}}{\left\{
\phi\left(  \zeta^{\prime}\right)  -\left[  \phi\left(  \zeta\right)  \right]
^{\ast}\right\}  \left\{  \left[  \phi\left(  \zeta^{\prime}\right)  \right]
^{\ast}-\phi\left(  \zeta\right)  \right\}  }.
\label{GF-polygon-GF-semiplane-5}
\end{equation}

\subsection{Limit 1}

Differentiating eq.~(\ref{GF-polygon-GF-semiplane-5}), we find
\begin{align}
\frac{\partial}{\partial\zeta^{\prime}}\left\langle \zeta^{\prime}\right|
\left[  \Delta\left(  C\right)  \right]  ^{-1}\left|  \zeta\right\rangle  &
=\frac{1}{4\pi}\frac{\partial}{\partial\zeta^{\prime}}\ln\frac{\phi\left(
\zeta^{\prime}\right)  -\phi\left(  \zeta\right)  }{\phi\left(  \zeta^{\prime
}\right)  -\left[  \phi\left(  \zeta\right)  \right]  ^{\ast}}\nonumber\\
&  =\frac{1}{4\pi}\left[  \frac{\frac{d}{d\zeta^{\prime}}\phi\left(
\zeta^{\prime}\right)  }{\phi\left(  \zeta^{\prime}\right)  -\phi\left(
\zeta\right)  }-\frac{\frac{d}{d\zeta^{\prime}}\phi\left(  \zeta^{\prime
}\right)  }{\phi\left(  \zeta^{\prime}\right)  -\left[  \phi\left(
\zeta\right)  \right]  ^{\ast}}\right]  . \label{diag-limit-calc-5}
\end{align}
The second differentiation gives
\begin{align}
\frac{\partial}{\partial\zeta}\frac{\partial}{\partial\zeta^{\prime}
}\left\langle \zeta^{\prime}\right|  \left[  \Delta\left(  C\right)  \right]
^{-1}\left|  \zeta\right\rangle  &  =\frac{1}{4\pi}\frac{\partial}
{\partial\zeta}\left[  \frac{\frac{d}{d\zeta^{\prime}}\phi\left(
\zeta^{\prime}\right)  }{\phi\left(  \zeta^{\prime}\right)  -\phi\left(
\zeta\right)  }\right] \nonumber\\
&  =\frac{1}{4\pi}\frac{\left[  \frac{d}{d\zeta}\phi\left(  \zeta\right)
\right]  \left[  \frac{d}{d\zeta^{\prime}}\phi\left(  \zeta^{\prime}\right)
\right]  }{\left[  \phi\left(  \zeta^{\prime}\right)  -\phi\left(
\zeta\right)  \right]  ^{2}}.
\end{align}
Next we set
\begin{equation}
\zeta^{\prime}=\zeta+\eta,
\end{equation}
\begin{equation}
\left[  \frac{\partial}{\partial\zeta}\frac{\partial}{\partial\zeta^{\prime}
}\left\langle \zeta^{\prime}\right|  \left[  \Delta\left(  C\right)  \right]
^{-1}\left|  \zeta\right\rangle \right]  _{\zeta^{\prime}=\zeta+\eta}
=\frac{1}{4\pi}\frac{\phi^{\prime}\left(  \zeta\right)  \phi^{\prime}\left(
\zeta+\eta\right)  }{\left[  \phi\left(  \zeta+\eta\right)  -\phi\left(
\zeta\right)  \right]  ^{2}}
\end{equation}
and expand the RHS in powers of $\eta$
\begin{align}
\frac{\phi^{\prime}\left(  \zeta\right)  \phi^{\prime}\left(  \zeta
+\eta\right)  }{\left[  \phi\left(  \zeta+\eta\right)  -\phi\left(
\zeta\right)  \right]  ^{2}}  &  =\frac{\left[  \phi^{\prime}\left(
\zeta\right)  \right]  ^{2}+\eta\phi^{\prime}\left(  \zeta\right)
\phi^{\prime\prime}\left(  \zeta\right)  +\frac{1}{2}\eta^{2}\phi^{\prime
}\left(  \zeta\right)  \phi^{\prime\prime\prime}\left(  \zeta\right)  \ldots
}{\eta^{2}\left[  \phi^{\prime}\left(  \zeta\right)  +\frac{1}{2}\eta
\phi^{\prime\prime}\left(  \zeta\right)  +\frac{1}{6}\eta^{2}\phi
^{\prime\prime\prime}\left(  \zeta\right)  +\ldots\right]  ^{2}}\nonumber\\
&  =\frac{1}{\eta^{2}}+\frac{1}{6}\left[  \frac{\phi^{\prime\prime\prime}
}{\phi^{\prime}}-\frac{3}{2}\left(  \frac{\phi^{\prime\prime}}{\phi^{\prime}
}\right)  ^{2}\right]  _{\zeta}+O\left(  \eta\right)  .
\end{align}
This can be expressed via Schwarz derivative (see eq.
(\ref{Schwarz-derivative-def-3}) in appendix~\ref{Appendix-Schwarz-derivative}
):
\begin{equation}
\{\phi,\zeta\}=\frac{\phi^{\prime\prime\prime}}{\phi^{\prime}}-\frac{3}
{2}\left(  \frac{\phi^{\prime\prime}}{\phi^{\prime}}\right)  ^{2}.
\end{equation}
Hence
\begin{equation}
\left[  \frac{\partial}{\partial\zeta}\frac{\partial}{\partial\zeta^{\prime}
}\left\langle \zeta^{\prime}\right|  \left[  \Delta\left(  C\right)  \right]
^{-1}\left|  \zeta\right\rangle \right]  _{\zeta^{\prime}=\zeta+\eta}
=\frac{1}{4\pi}\left(  \frac{1}{\eta^{2}}+\frac{1}{6}\{\phi,\zeta\}\right)
+O\left(  \eta\right)  .
\end{equation}
Since $\phi$ describes conformal mapping $z=\phi\left(  \zeta\right)  $, we
can identify
\begin{equation}
\{\phi,\zeta\}=\{z,\zeta\}
\end{equation}
so that
\begin{equation}
\lim_{\eta\rightarrow0}\left\{  \left[  \frac{\partial}{\partial\zeta
}\frac{\partial}{\partial\zeta^{\prime}}\left\langle \zeta^{\prime}\right|
\left[  \Delta\left(  C\right)  \right]  ^{-1}\left|  \zeta\right\rangle
\right]  _{\zeta^{\prime}=\zeta+\eta}-\frac{1}{4\pi}\frac{1}{\eta^{2}
}\right\}  =\frac{1}{24\pi}\{z,\zeta\}.
\end{equation}
Combining this with eq.~(\ref{K-zet-zeta-def}), we obtain eq.
(\ref{K-zeta-zeta-CS}).

\subsection{Limit 2}

Applying derivative $\partial/\partial\zeta^{\ast}$ to eq.
(\ref{diag-limit-calc-5}), we find
\begin{align}
\frac{\partial}{\partial\zeta^{\ast}}\frac{\partial}{\partial\zeta^{\prime}
}\left\langle \zeta^{\prime}\right|  \left[  \Delta\left(  C\right)  \right]
^{-1}\left|  \zeta\right\rangle  &  =\frac{1}{4\pi}\frac{\partial}
{\partial\zeta^{\ast}}\left[  -\frac{\frac{d}{d\zeta^{\prime}}\phi\left(
\zeta^{\prime}\right)  }{\phi\left(  \zeta^{\prime}\right)  -\left[
\phi\left(  \zeta\right)  \right]  ^{\ast}}\right] \nonumber\\
&  =-\frac{1}{4\pi}\frac{\frac{d}{d\zeta^{\prime}}\phi\left(  \zeta^{\prime
}\right)  \left[  \frac{d}{d\zeta^{\prime}}\phi\left(  \zeta\right)  \right]
^{\ast}}{\left[  \phi\left(  \zeta\right)  -\left[  \phi\left(  \zeta\right)
\right]  ^{\ast}\right]  ^{2}}.
\end{align}
Taking the limit $\zeta^{\prime}\rightarrow\zeta$, we obtain
\begin{align}
\lim_{\zeta^{\prime}\rightarrow\zeta}\left[  \frac{\partial}{\partial
\zeta^{\ast}}\frac{\partial}{\partial\zeta^{\prime}}\left\langle \zeta
^{\prime}\right|  \left[  \Delta\left(  C\right)  \right]  ^{-1}\left|
\zeta\right\rangle \right]   &  =-\frac{1}{4\pi}\frac{\left|  \phi^{\prime
}\left(  \zeta\right)  \right|  ^{2}}{\left[  \phi\left(  \zeta\right)
-\left[  \phi\left(  \zeta\right)  \right]  ^{\ast}\right]  ^{2}}\nonumber\\
&  =-\frac{1}{4\pi}\left|  \frac{dz}{d\zeta}\right|  ^{2}\frac{1}{\left(
z-z^{\ast}\right)  ^{2}}=\frac{1}{16\pi}\left|  \frac{dz}{d\zeta}\right|
^{2}\left|  \operatorname{Im}z\right|  ^{-2}.
\end{align}
Combining this with eq.~(\ref{K-zeta-zeta-bar-def}), we derive eq.
(\ref{K-zeta-zeta-bar-CS}).

\section{SC mapping for triangles}

\setcounter{equation}{0} 

\label{Appendix-SC-mapping-triangles}

In this appendix we apply SC equation (\ref{SC-tilded-1}) with one vertex at
infinity to the case of triangles and derive some useful relations between
geometric parameters and SC\ parameters. The case of triangles corresponds to
setting $n_{\text{v}}=3$ in eq.~(\ref{SC-tilded-1}):
\begin{equation}
\frac{d\zeta}{dz}=\tilde{A}\prod_{k=1}^{2}\left(  z-z_{k}\right)  ^{-\beta
_{k}}. \label{SC-mapping-triangle}
\end{equation}
Choosing
\begin{equation}
z_{1}=0,\quad z_{2}=1, \label{z1-z2-standard}
\end{equation}
we simplify SC\ equation (\ref{SC-mapping-triangle}) to
\begin{equation}
\frac{d\zeta}{dz}=\tilde{A}z^{-\beta_{1}}\left(  z-1\right)  ^{-\beta_{2}}.
\end{equation}
The length of the triangle side SC-mapped to real interval
\begin{equation}
\lbrack z_{1},z_{2}]=\left[  0,1\right]
\end{equation}
equals
\begin{align}
L_{12}  &  =\int_{0}^{1}dz\left|  \frac{d\zeta}{dz}\right|  =\left|
\tilde{A}\right|  \int_{0}^{1}dz\,z^{-\beta_{1}}\left(  1-z\right)
^{-\beta_{2}}\nonumber\\
&  =\left|  \tilde{A}\right|  B\left(  1-\beta_{1},1-\beta_{2}\right)
=\left|  \tilde{A}\right|  \frac{\Gamma\left(  1-\beta_{1}\right)
\Gamma\left(  1-\beta_{2}\right)  }{\Gamma\left(  2-\beta_{1}-\beta
_{2}\right)  }.
\end{align}
With $\beta_{3}$ given by eq.~(\ref{beta-n-v-via-other})
\begin{equation}
\beta_{3}=2-\beta_{1}-\beta_{2}\, \label{beta-3-via-beta-1-beta-2}
\end{equation}
we find
\begin{equation}
L_{12}=\left|  \tilde{A}\right|  \frac{\Gamma\left(  1-\beta_{1}\right)
\Gamma\left(  1-\beta_{2}\right)  }{\Gamma\left(  \beta_{3}\right)  }.
\end{equation}
Area $S\left(  C\right)  $ of the triangle can be computed using elementary
geometry:
\begin{equation}
S\left(  C\right)  =\frac{1}{2}L_{12}L_{13}\sin\theta_{1},
\end{equation}
\begin{equation}
\frac{L_{12}}{L_{13}}=\frac{\sin\theta_{3}}{\sin\theta_{2}}.
\end{equation}
Here\ $L_{13}$ is the length of the triangle side opposite to the vertex with
angle $\theta_{2}$. Angles $\theta_{k}$ of the triangle are given by eq.
(\ref{theta-k-via-beta-k}). Now we have
\begin{align}
S\left(  C\right)   &  =\frac{1}{2}L_{12}^{2}\frac{\sin\theta_{1}\sin
\theta_{2}}{\sin\theta_{3}}\nonumber\\
&  =\frac{1}{2}\left[  \left|  \tilde{A}\right|  \frac{\Gamma\left(
1-\beta_{1}\right)  \Gamma\left(  1-\beta_{2}\right)  }{\Gamma\left(
\beta_{3}\right)  }\right]  ^{2}\frac{\sin\pi\left(  1-\beta_{1}\right)
\sin\pi\left(  1-\beta_{2}\right)  }{\sin\pi\left(  1-\beta_{3}\right)  },
\end{align}
which can be simplified to
\begin{align}
\left|  \tilde{A}\right|  ^{-2}  &  =\frac{\pi}{2}\frac{1}{S\left(  C\right)
}\frac{\Gamma\left(  1-\beta_{1}\right)  \Gamma\left(  1-\beta_{2}\right)
\Gamma\left(  1-\beta_{3}\right)  }{\Gamma\left(  \beta_{1}\right)
\Gamma\left(  \beta_{2}\right)  \Gamma\left(  \beta_{3}\right)  }
\,\label{A-tilde-via-S-infty}\\
&  \left(  z_{1}=0,\,z_{2}=1\right)  .
\end{align}
This result can be easily generalized to the case of arbitrary $z_{1},z_{2}$:

\begin{equation}
S\left(  C\right)  =\frac{\pi}{2}\left[  \left|  \tilde{A}\right|  \left|
z_{1}-z_{2}\right|  ^{-\beta_{1}-\beta_{2}+1}\right]  ^{2}\frac{\Gamma\left(
1-\beta_{1}\right)  \Gamma\left(  1-\beta_{2}\right)  \Gamma\left(
1-\beta_{3}\right)  }{\Gamma\left(  \beta_{1}\right)  \Gamma\left(  \beta
_{2}\right)  \Gamma\left(  \beta_{3}\right)  }. \label{S-via-A-infty-2}
\end{equation}

\section{Calculation of $\Pi_{1}^{\left(  2\right)  }$}

\setcounter{equation}{0} 

\label{Appendix-K-1-2-calculation}

Setting
\begin{equation}
n=2,\quad z_{1}=0,\quad z=x+iy
\end{equation}
in eq.~(\ref{K-P-def-integral-naive-1}), we find
\begin{align}
&  \Pi_{1}^{(2)\text{ }}\left(  \alpha,\left\{  \gamma_{1},\gamma_{2}\right\}
,\left\{  0,z_{2}\right\}  \right) \nonumber\\
&  =\int_{\mathbb{C}_{+}}d^{2}z\,\left|  \operatorname{Im}\,z\right|
^{\alpha-1}\left|  z\right|  ^{\gamma_{1}-\alpha-1}\left|  z-z_{2}\right|
^{\gamma_{2}-\alpha-1}\nonumber\\
&  =\int_{-\infty}^{\infty}dx\int_{0}^{\infty}dy\,y^{\alpha-1}\left|
x^{2}+y^{2}\right|  ^{\left(  \gamma_{1}-\alpha-1\right)  /2}\left|  \left(
x-z_{2}\right)  ^{2}+y^{2}\right|  ^{\left(  \gamma_{2}-\alpha-1\right)  /2}.
\label{Pi-2-1-int-def}
\end{align}
Integrals of this type appear in various problems of mathematical physics.
In appendix B of ref.~\cite{BB-2010} this integral was computed
(in a slightly different notation assuming $z_2=1$) using
advanced techniques including a reduction
to higher hypergeometric function ${}_3F_2$. Below an
elementary calculation of integral (\ref{Pi-2-1-int-def}) is sketched.

Using the representation
\begin{equation}
\frac{1}{D_{1}^{\alpha_{1}}D_{2}^{\alpha_{2}}}=\frac{\Gamma(\alpha_{1}
+\alpha_{2})}{\Gamma(\alpha_{1})\Gamma(\alpha_{2})}\int\limits_{0}^{1}
d\mu\frac{\mu^{\alpha_{1}-1}\left(  1-\mu\right)  ^{\alpha_{2}-1}}{\left[
(D_{1}\mu+D_{2}\left(  1-\mu\right)  \right]  ^{\alpha_{1}+\alpha_{2}}}
\label{Feynman-trick}
\end{equation}
for the factor
\begin{align}
&  \left|  x^{2}+y^{2}\right|  ^{\left(  \gamma_{1}-\alpha-1\right)
/2}\left|  \left(  x-z_{2}\right)  ^{2}+y^{2}\right|  ^{\left(  \gamma
_{2}-\alpha-1\right)  /2}\nonumber\\
&  =\frac{\Gamma\left(  \alpha+1-\frac{\gamma_{1}+\gamma_{2}}{2}\right)
}{\Gamma\left(  \frac{\alpha+1-\gamma_{1}}{2}\right)  \Gamma\left(
\frac{\alpha+1-\gamma_{1}}{2}\right)  }\int\limits_{0}^{1}d\mu\,\left(
1-\mu\right)  ^{\frac{\alpha-1-\gamma_{1}}{2}}\,\mu^{\frac{\alpha-1-\gamma
_{2}}{2}}\nonumber\\
&  \times\left\{  \left[  \left(  1-\mu\right)  \left(  x^{2}+y^{2}\right)
+\mu\left[  \left(  x-z_{2}\right)  ^{2}+y^{2}\right]  \right]  \right\}
^{\frac{\gamma_{1}+\gamma_{2}}{2}-\alpha-1}
\end{align}
and changing the order of integrations, we obtain
\begin{align}
&  \Pi_{1}^{(2)}\left(  \alpha,\left\{  \gamma_{1},\gamma_{2}\right\}
,\left\{  0,z_{2}\right\}  \right)  =\frac{\Gamma\left(  \alpha+1-\frac{\gamma
_{1}+\gamma_{2}}{2}\right)  }{\Gamma\left(  \frac{\alpha+1-\gamma_{1}}
{2}\right)  \Gamma\left(  \frac{\alpha+1-\gamma_{2}}{2}\right)  }\nonumber\\
&  \times\int\limits_{0}^{1}d\mu\,\left(  1-\mu\right)  ^{\frac{\alpha
-1-\gamma_{1}}{2}}\,\mu^{\frac{\alpha-1-\gamma_{2}}{2}}\int_{0}^{\infty
}dy\,y^{\alpha-1}\int_{-\infty}^{\infty}dx\nonumber\\
&  \times\left\{  \left[  \left(  1-\mu\right)  \left(  x^{2}+y^{2}\right)
+\mu\left[  \left(  x-z_{2}\right)  ^{2}+y^{2}\right]  \right]  \right\}
^{\frac{\gamma_{1}+\gamma_{2}}{2}-\alpha-1}. \label{K-1-2-calc-2}
\end{align}
This order of integrations reduces the calculation to a chain of simple
B-function integrals resulting in
\[
\int\limits_{0}^{1}d\mu\,\left(  1-\mu\right)  ^{\frac{\alpha-1-\gamma_{1}}
{2}}\,\mu^{\frac{\alpha-1-\gamma_{2}}{2}}\int_{0}^{\infty}dy\,y^{\alpha-1}
\int_{-\infty}^{\infty}dx
\]
\begin{align}
&  \times\left\{  \left[  \left(  1-\mu\right)  \left(  x^{2}+y^{2}\right)
+\mu\left[  \left(  x-z_{2}\right)  ^{2}+y^{2}\right]  \right]  \right\}
^{\frac{\gamma_{1}+\gamma_{2}}{2}-\alpha-1}\nonumber\\
&  =\frac{\sqrt{\pi}}{2}\left|  z_{2}\right|  ^{\gamma_{1}+\gamma_{2}
-1-\alpha}\Gamma\left(  \frac{\alpha}{2}\right)  \frac{\Gamma\left(
\frac{\alpha+1}{2}-\frac{\gamma_{1}+\gamma_{2}}{2}\right)  }{\Gamma\left(
\alpha+1-\frac{\gamma_{1}+\gamma_{2}}{2}\right)  }\frac{\Gamma\left(
\frac{\gamma_{1}}{2}\right)  \Gamma\left(  \frac{\gamma_{2}}{2}\right)
}{\Gamma\left(  \frac{\gamma_{1}+\gamma_{2}}{2}\right)  }.
\end{align}

We insert this into eq.~(\ref{K-1-2-calc-2})
\begin{align}
&  \Pi_{1}^{(2)}\left(  \alpha,\left\{  \gamma_{1},\gamma_{2}\right\}
,\left\{  0,z\right\}  \right) \nonumber\\
&  =\frac{\sqrt{\pi}}{2}\left(  z_{2}\right)  ^{\gamma_{1}+\gamma_{2}
-1-\alpha}\Gamma\left(  \frac{\alpha}{2}\right)  \frac{\Gamma\left(
\frac{\gamma_{1}}{2}\right)  }{\Gamma\left(  \frac{\alpha+1}{2}-\frac{\gamma
_{1}}{2}\right)  }\frac{\Gamma\left(  \frac{\gamma_{2}}{2}\right)  }
{\Gamma\left(  \frac{\alpha+1}{2}-\frac{\gamma_{2}}{2}\right)  }
\frac{\Gamma\left(  \frac{\alpha+1}{2}-\frac{\gamma_{1}+\gamma_{2}}{2}\right)
}{\Gamma\left(  \frac{\gamma_{1}+\gamma_{2}}{2}\right)  }.
\end{align}
Introducing notation
\begin{equation}
\gamma_{3}=\left(  \alpha+1\right)  -\left(  \gamma_{1}+\gamma_{2}\right)  ,
\end{equation}
we find
\begin{align}
&  \Pi_{1}^{(2)\text{ }}\left(  \alpha,\left\{  \gamma_{1},\gamma_{2}\right\}
,\left\{  0,z_{2}\right\}  \right) \nonumber\\
&  =\left[  \frac{\sqrt{\pi}}{2}\left|  z_{2}\right|  ^{-\gamma_{3}}
\Gamma\left(  \frac{\alpha}{2}\right)  \prod_{k=1}^{3}\frac{\Gamma\left(
\frac{\gamma_{k}}{2}\right)  }{\Gamma\left(  \frac{\alpha+1}{2}-\frac{\gamma
_{k}}{2}\right)  }\right]  _{\gamma_{3}=\left(  \alpha+1\right)  -\left(
\gamma_{1}+\gamma_{2}\right)  }. \label{K-1-2-res-0}
\end{align}
This result was derived in the region of parameters
$\alpha$, $\gamma_1$, $\gamma_2$ where integral (\ref{Pi-2-1-int-def})
and other intermediate integrals are convergent.
The RHS is a meromorphic
function of $\alpha,\gamma_{1},\gamma_{2}$ in agreement with the general
properties of functions $\Pi_{P}^{(n)\text{ }}$ discussed
in section~\ref{functions-Pi-section}.

Using translational invariance, we generalize eq.~(\ref{K-1-2-res-0}) to
\begin{align}
&  \Pi_{1}^{(2)}\left(  \alpha,\left\{  \gamma_{1},\gamma_{2}\right\}
,\left\{  z_{1},z_{2}\right\}  \right) \nonumber\\
&  =\left[  \frac{\sqrt{\pi}}{2}\left|  z_{1}-z_{2}\right|  ^{-\gamma_{3}
}\Gamma\left(  \frac{\alpha}{2}\right)  \prod_{k=1}^{3}\frac{\Gamma\left(
\frac{\gamma_{k}}{2}\right)  }{\Gamma\left(  \frac{\alpha+1}{2}-\frac{\gamma
_{k}}{2}\right)  }\right]  _{\gamma_{3}=\left(  \alpha+1\right)  -\left(
\gamma_{1}+\gamma_{2}\right)  }. \label{K-1-2-res}
\end{align}

\section{Calculation of $\Pi_{P}^{(2)}$}

\label{Appendix-K-P-2-calculation} \setcounter{equation}{0} 

In this appendix we compute function $\Pi_{P}^{(2)}\left(  \alpha,\left\{
\gamma_{1},\gamma_{2}\right\}  ,\left\{  z_{1},z_{2}\right\}  \right)  $ for
the special value of argument $\alpha=1$. For simplicity we choose $z_{1}=0$,
$z_{2}=1$. We will perform a formal calculation ignoring divergences of some
intermediate integrals but still leading to the correct final result. A
rigorous justification of this formal calculation requires methods whose
discussion makes sense in the context of the theory of functions $\Pi
_{P}^{(n)}$ with arbitrary $n$, i.e. in the case of the theory of analytical
regularization for arbitrary polygons $C$.

We start from the integral over complex plane $\mathbb{C}$
\begin{equation}
\int_{\mathbb{C}}d^{2}z\,\left|  z\right|  ^{2A}\left|  z-w\right|
^{2B}=\pi\left|  w\right|  ^{2A+2B+2}\frac{\Gamma\left(  A+1\right)
\Gamma\left(  B+1\right)  \Gamma\left(  -A-B-1\right)  }{\Gamma\left(
-A\right)  \Gamma\left(  -B\right)  \Gamma\left(  A+B+2\right)  }
\label{int-z-1}
\end{equation}
which can be easily computed (e.g., using representation (\ref{Feynman-trick})).
Applying derivatives
\begin{equation}
\left(  \frac{\partial}{\partial w}\right)  ^{m}\left(  \frac{\partial
}{\partial w^{\ast}}\right)  ^{n}
\end{equation}
to identity (\ref{int-z-1}), we find
\begin{align}
&  \left(  -1\right)  ^{m+n}\frac{\Gamma\left(  B+1\right)  }{\Gamma\left(
B+1-m\right)  }\frac{\Gamma\left(  B+1\right)  }{\Gamma\left(  B+1-n\right)
}\int_{\mathbb{C}}d^{2}z\,\frac{\left|  z\right|  ^{2A}\left|  z-w\right|
^{2B}}{\left(  z-w\right)  ^{m}\left(  z^{\ast}-w^{\ast}\right)  ^{n}
}\nonumber\\
&  =\pi\frac{\Gamma\left(  A+1\right)  \Gamma\left(  B+1\right)  \Gamma\left(
-A-B-1\right)  }{\Gamma\left(  -A\right)  \Gamma\left(  -B\right)
\Gamma\left(  A+B+2\right)  }\frac{\Gamma\left(  A+B+2\right)  }{\Gamma\left(
A+B+2-m\right)  }\frac{\Gamma\left(  A+B+2\right)  }{\Gamma\left(
A+B+2-n\right)  }\nonumber\\
&  \times w^{A+B+1-m}\left(  w^{\ast}\right)  ^{A+B+1-n}.
\end{align}
Replacing integration variable $z\rightarrow z-w$ and exchanging
$A\leftrightarrow B$, we obtain after simplifications
\begin{align}
&  \int_{\mathbb{C}}d^{2}z\,\left|  z\right|  ^{2A}\left|  z-w\right|
^{2B}z^{m}\left(  z^{\ast}\right)  ^{n}\nonumber\\
&  =-\pi\frac{\sin\left(  \pi A\right)  }{\sin\left[  \pi\left(  A+B\right)
\right]  }\frac{\Gamma\left(  B+1\right)  }{\Gamma\left(  -B\right)
}\frac{\Gamma\left(  A+1+m\right)  \Gamma\left(  A+1+n\right)  }{\Gamma\left(
A+B+2+m\right)  \Gamma\left(  A+B+2+n\right)  }\nonumber\\
&  \times\left|  w\right|  ^{2\left(  A+B+1\right)  }w^{m}\left(  w^{\ast
}\right)  ^{n}. \label{int-z-w-m-n}
\end{align}

Next we set $w=1$ and notice that after the symmetrization $m\leftrightarrow
n$ in eq.~(\ref{int-z-w-m-n}) one can reduce the integral over complex plane $\mathbb{C}$
to the integral over upper semiplane $\mathbb{C}_{+}$
(\ref{C-2-plus-def}):
\begin{align}
&  \int_{\mathbb{C}}d^{2}z\,\left|  z\right|  ^{2A}\left|  z-1\right|
^{2B}\left[  z^{m}\left(  z^{\ast}\right)  ^{n}+z^{n}\left(  z^{\ast}\right)
^{m}\right]  \nonumber\\
&  =2\int_{\mathbb{C}_{+}}d^{2}z\,\left|  z\right|  ^{2A}\left|
z-1\right|  ^{2B}\left[  z^{m}\left(  z^{\ast}\right)  ^{n}+z^{n}\left(
z^{\ast}\right)  ^{m}\right]  .
\end{align}
Hence
\begin{align}
\frac{1}{2} &  \int_{\mathbb{C}_{+}}d^{2}z\,\left|  z\right|  ^{2A}\left|
z-1\right|  ^{2B}\left[  z^{m}\left(  z^{\ast}\right)  ^{n}+z^{n}\left(
z^{\ast}\right)  ^{m}\right]  \nonumber\\
&  =-\frac{\pi}{2}\frac{\sin\left(  \pi A\right)  }{\sin\left[  \pi\left(
A+B\right)  \right]  }\frac{\Gamma\left(  B+1\right)  }{\Gamma\left(
-B\right)  }\frac{\Gamma\left(  A+1+m\right)  \Gamma\left(  A+1+n\right)
}{\Gamma\left(  A+B+2+m\right)  \Gamma\left(  A+B+2+n\right)  }.
\label{int-for-Pi-2-P-m-n}
\end{align}
The LHS can be interpreted as integral (\ref{K-P-def-integral-naive-1}) defining function $\Pi_{P}^{(2)}$
with a special choice of its polynomial argument $P$:
\begin{equation}
P_{mn}\left(  z\right)  =\frac{1}{2}\left[  z^{m}\left(  z^{\ast}\right)
^{n}\,+z^{n}\left(  z^{\ast}\right)  ^{m}\right]  .
\end{equation}
Setting
\begin{equation}
n=2,\,\alpha=1,
\end{equation}
\begin{equation}
z_{1}=0,\,z_{2}=1,
\end{equation}
in eq.~(\ref{K-P-def-integral-naive-1}), we find from eq.~(\ref{int-for-Pi-2-P-m-n})

\begin{align}
\Pi_{P_{mn}}^{(2)\text{ }}\left(  1,\left\{  \gamma_{1},\gamma_{2}\right\}
,\left\{  0,1\right\}  \right)   &  =\frac{\pi}{2}\frac{\sin\left(
\pi\frac{\gamma_{1}}{2}\right)  }{\sin\left(  \pi\frac{\gamma_{1}+\gamma_{2}
}{2}\right)  }\frac{\Gamma\left(  \frac{\gamma_{2}}{2}\right)  }{\Gamma\left(
1-\frac{\gamma_{2}}{2}\right)  }\nonumber\\
&  \times\frac{\Gamma\left(  \frac{\gamma_{1}}{2}+m\right)  \Gamma\left(
\frac{\gamma_{1}}{2}+n\right)  }{\Gamma\left(  \frac{\gamma_{1}+\gamma_{2}}
{2}+m\right)  \Gamma\left(  \frac{\gamma_{1}+\gamma_{2}}{2}+n\right)  }.
\label{K-P-mn-2-res}
\end{align}

For a general polynomial
\begin{equation}
P\left(  z,z^{\ast}\right)  =\sum_{mn}c_{mn}z^{m}\left(  z^{\ast}\right)  ^{n}
\end{equation}
with symmetric coefficients
\begin{equation}
c_{mn}=c_{nm}
\end{equation}
we derive from eq.~(\ref{K-P-mn-2-res}):
\begin{align}
\Pi_{P}^{(2)\text{ }}\left(  1,\left\{  \gamma_{1},\gamma_{2}\right\}
,\left\{  0,1\right\}  \right)   &  =\frac{\pi}{2}\frac{\sin\left(
\pi\frac{\gamma_{1}}{2}\right)  }{\sin\left(  \pi\frac{\gamma_{1}+\gamma_{2}
}{2}\right)  }\frac{\Gamma\left(  \frac{\gamma_{2}}{2}\right)  }{\Gamma\left(
1-\frac{\gamma_{2}}{2}\right)  }\nonumber\\
&  \times\sum_{mn}c_{mn}\frac{\Gamma\left(  \frac{\gamma_{1}}{2}+m\right)
\Gamma\left(  \frac{\gamma_{1}}{2}+n\right)  }{\Gamma\left(  \frac{\gamma
_{1}+\gamma_{2}}{2}+m\right)  \Gamma\left(  \frac{\gamma_{1}+\gamma_{2}}
{2}+n\right)  }. \label{K-P-2-generl-res}
\end{align}

\end{document}